\documentclass[9pt,twocolumn,twoside]{pnas-new}

\templatetype{pnasresearcharticle} 

\usepackage{wrapfig}
\usepackage{graphicx}
\usepackage{amsmath}

\newcommand{\resub}[1]{\textcolor{black}{#1}}
\newcommand{\rrs}[1]{\textcolor{black}{#1}}  
\newcommand{\fs}[1]{\textcolor{black}{#1}}  

\newcommand{\um}{~\mu\mathrm{m}}
\newcommand{\rs}{~\mathrm{rad}~\mathrm{s}^{-1}}
\newcommand{\MM}{\textit{Materials and Methods}}
\newcommand{\SI}{\textit{Supplementary Information}}

\title{
Enhanced microscopic dynamics in mucus gels under a mechanical load in the linear \rrs{viscoelastic} regime}

\author[a,1,2]{Domenico Larobina}
\author[b,c]{Angelo Pommella}
\author[b,d]{Adrian-Marie Philippe}
\author[b,e]{Med Yassine Nagazi}
\author[b,f,1,2]{Luca Cipelletti}

\affil[a]{Institute for Polymers, Composites and Biomaterials, National Research Council of Italy, P.le E. Fermi 1, Naples, 80055 Portici, Italy}
\affil[b]{Laboratoire Charles Coulomb (L2C), Universit\'e Montpellier, CNRS, Montpellier, France}
\affil[c]{Present address: Univ Lyon, INSA Lyon, CNRS, MATEIS, UMR5510, F-69621 Villeurbanne, France}
\affil[d]{Present address: Materials Research and Technology Department, Luxembourg Institute of Science and Technology, 41 rue du Brill, L-4422 Belvaux, Luxembourg}
\affil[e]{Present address: Formulaction, 31200 Toulouse, France}
\affil[f]{Institut Universitaire de France (IUF), France}

\leadauthor{Lead author last name}

\significancestatement{Mucus is a biological gel protecting several tissues. Its key properties result from a crucial balance between solid-like and fluid-like behavior, insured by the non-permanent nature of the bonds between its macromolecular constituents. Our understanding of the micron-scale response of mucus to an applied stress is still rudimentary, although in living organisms stresses acting on mucus are ubiquitous, from bacterial penetration to coughing and peristalsis. We show that under a modest applied stress, in the mechanical linear regime, the microscopic dynamics of pig gastric mucus transiently accelerate by up to two orders of magnitude. A simple model rationalizes this previously unrecognized fluidization mechanism stemming from elastic recoil following bond breaking and generalizes our findings to networks with reversible bonds.}

\authorcontributions{DL, and LC designed research and elaborated the model; DL, AP, AMP, LC performed research. DL, AP, LC analyzed data; YN contributed to the model; DL and LC wrote the paper.}
\authordeclaration{The authors declare no conflict of interest.}
\equalauthors{\textsuperscript{1}D.L and L.C.contributed equally to this work.}
\correspondingauthor{\textsuperscript{2}To whom correspondence should be addressed. E-mail: domenico.larobina@cnr.it; luca.cipelletti@umontpellier.fr}

\keywords{Mucus $|$ Rheology $|$ Dynamic Light Scattering $|$ Stress relaxation $|$ Microscopic dynamics}

\begin{abstract}
Mucus is a biological gel covering the surface of several tissues and insuring key biological functions, including as a protective barrier against dehydration, pathogens penetration, or gastric acids. Mucus biological functioning requires a finely tuned balance between solid-like and fluid-like mechanical response, insured by reversible bonds between mucins, the glycoproteins that form the gel. In living organisms, mucus is subject to various kinds of mechanical stresses, e.g. due to osmosis, bacterial penetration, coughing and gastric peristalsis. However, our knowledge of the effects of stress on mucus is still rudimentary and mostly limited to macroscopic rheological measurements, with no insight into the relevant microscopic mechanisms. Here, we run mechanical tests simultaneously to measurements of the microscopic dynamics of pig gastric mucus. Strikingly, we find that a modest shear stress, within the macroscopic rheological linear regime, dramatically enhances mucus reorganization at the microscopic level, as signalled by a transient acceleration of the microscopic dynamics, by up to two orders of magnitude. We rationalize these findings by proposing a simple yet general model for the dynamics of physical gels under strain and validate its assumptions through numerical simulations of spring networks. These results shed new light on the rearrangement dynamics of mucus at the microscopic scale, with potential implications in phenomena ranging from mucus clearance to bacterial and drug penetration.
\end{abstract}

\dates{This manuscript was compiled on \today}
\doi{\url{www.pnas.org/cgi/doi/10.1073/pnas.XXXXXXXXXX}}

\begin{document}

\maketitle
\thispagestyle{firststyle}
\ifthenelse{\boolean{shortarticle}}{\ifthenelse{\boolean{singlecolumn}}{\abscontentformatted}{\abscontent}}{}

\dropcap{M}ucus is a biogel ubiquitous across both vertebrates and invertebrates~\cite{denny_invertebrate_1989,vasquez2015complex,cone_barrier_2009}. The main mucus macromolecular components are a family of glycosylated proteins called mucins \cite{bansil_mucin_2006,wagner_mucins_2018,ridley2018mucins}. Hydrophobic, hydrogen bonding and $Ca^{2+}$-mediated \cite{meldrum_mucin_2018-1} interactions between mucins are responsible for macromolecular associations determining the viscoelastic properties of mucus, which in turn control its biological functions \cite{wagner_mucins_2018,vasquez2015complex}. Alteration of the  viscoelastic properties compromise mucus functionality, resulting in severe diseases \cite{lai_micro-_2009,cohn2006mucus}.

Mucus viscoelasticity stems from the reversible nature of the bonds between its constituents, which insure solid-like behavior on short time scales while allowing flow on longer time scales. Rheological studies on mucus reporting the frequency dependence of the storage, $G'$, and loss, $G''$, components of the dynamic modulus reveal $G' > G''$, with $G'$ only weakly dependent on \fs{angular frequency $\omega$} on time scales 0.1-100 s \cite{cohn2006mucus,lai_micro-_2009,georgiades_particle_2014}, a behavior typical of soft solids \cite{stokes2008rheology}. Stress relaxation tests probe viscoelasticity on longer time scales, up to thousands of seconds. They reveal a power law or logarithmic decay of the shear stress with time~\cite{denny1980physical,denny1983molecular,Philippe2017}, indicative of a wide distribution of relaxation times, ascribed to the variety of macromolecular association mechanisms and the mucus complex, multiscale structure~\cite{Philippe2017,macierzanka2014transport,meldrum_mucin_2018-1}.

Alongside conventional rheology, microrheology has gained momentum, since it investigates the mechanical response of mucus on the length scales relevant to its biological functions, from a fraction of a micrcon up to $\sim 10 \um$~\resub{\cite{lai_micro-_2009,weigand_active_2017,wagner_rheological_2017,meldrum_mucin_2018-1,demouveaux_gel-forming_2018,jory_mucus_2019}}. Microrheology infers the viscoelastic moduli from the microscopic dynamics of tracer particles embedded in the sample~\cite{mason_optical_1995}, either due to spontaneous thermal fluctuations or externally driven, e.g., by a magnetic field.
\resub{Mucus viscoelasticity as measured by microrheology is found to depend on the size of the tracer particles, the local environment they probe, and the length scale over which their motion is tracked~\cite{lai_micro-_2009,weigand_active_2017,wagner_rheological_2017,meldrum_mucin_2018-1,jory_mucus_2019,bajka2015influence}}. Below $\approx 1 \um$, microrheology data are dominated by the diffusion of the probe particles within the mucus pores, \rrs{as inferred from the analysis of the localization of the tracers trajectories~\cite{meldrum_mucin_2018-1,wagner_rheological_2017}, their dependence on probe size~\cite{lai_micro-_2009,bajka2015influence,weigand_active_2017}, or on the amplitude of the external drive in active microrheology~\cite{weigand_active_2017}}. On larger length scales, microrheology reports the local viscoelasticity, which converges towards the macroscopic one above $\approx 10 \um$, \rrs{as revealed by the probe size and drive amplitude dependence of active microrheology}~\cite{weigand_active_2017}.

\textit{In vivo} mucus is submitted to stresses of various origin, involving strain on the microscopic scale as in cilia beating in muco-ciliary clearance~\cite{norton2011model} and bacterial penetration ~\cite{figueroa2019mechanical,celli_helicobacter_2009}, up to macroscopic scales, e.g. during coughing and peristaltis~\cite{king1987role,cone_barrier_2009}. \resub{Stresses due to the osmotic pressure exerted by the environment~\cite{datta_polymers_2016} or resulting from changes in hydration~\cite{button_periciliary_2012-1,anderson_relationship_2015} can modify the structure of mucus and, e.g.,} impair mucus clearance. By contrast, little is known on the impact of stress on the \textit{dynamics} of mucus, in particular at the microscopic level. Conventional rheology indicates that mucus is fluidized upon applying a large stress~\cite{ewoldt_rheological_2007,hyun2011review}, beyond the linear regime. This behavior is typical of soft solids~\cite{coblas2016correlation,bonn_yield_2017,donley_elucidating_2020}; in concentrated nanoemulsions and colloidal suspensions and in colloidal gels fluidization in the non-linear regime has been shown to stem from enhanced microscopic dynamics~\cite{viasnoff_rejuvenation_2002,schall_structural_2007,rogers_echoes_2014,sentjabrskaja_creep_2015,2020SciPy-NMeth,aime_microscopic_2018,rogers_microscopic_2018}. However, for mucus we still lack knowledge of the effect of an applied stress on the microscopic dynamics.

Here, we couple rheology and light and X-photon correlation methods to investigate the microscopic dynamics of pig gastric mucus under an applied shear stress. Surprisingly, we find that small stresses, well within the macroscopic linear viscoelastic regime, transiently enhance the mucus dynamics by up to two orders of magnitude. We propose a simple yet general model for the dynamics of physical gels under strain that rationalizes these findings.

\section*{Results}
\subsection*{Range of linear viscoelasticity}
\begin{figure}
    \centering
    \includegraphics{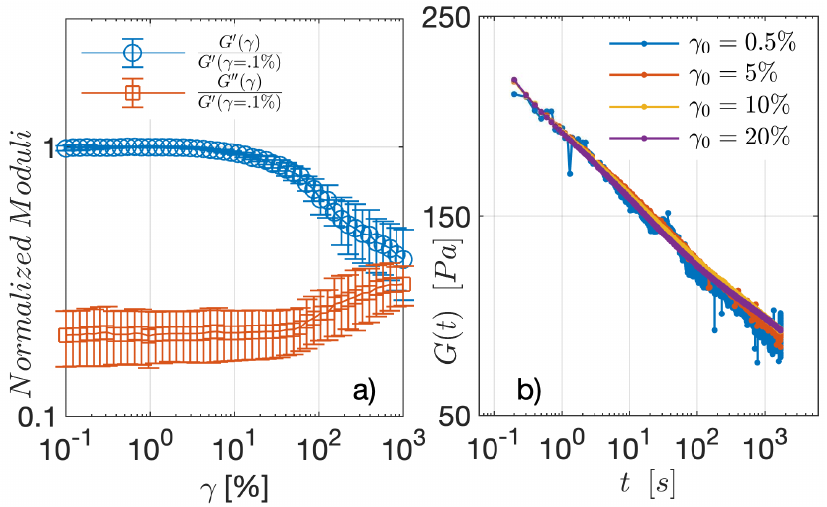}
    \caption{\textbf{Viscoelastic properties of pig gastric mucus gels}. (a) Normalized storage (circles) and loss (squares) moduli \textit{vs} applied strain in oscillatory shear rheology tests at a frequency \fs{$\omega = 6.28~\rs$}. Symbols; average over eight samples of the moduli normalized by $G'$ at the smallest strain. Error bars: standard deviation over the set of probed samples. b) Relaxation modulus following a step strain increment of amplitude $\gamma_0$, demonstrating linear behavior up to $20\%$.}
    \label{fig:Rheology}
\end{figure}
We measure the viscoelastic properties of mucus gels under shear and use oscillatory rheometry and stress relaxation tests to determine the range \rrs{of the operational linear viscoelastic regime, where the viscoelastic moduli do not depend on the applied strain or stress}. Figure~\ref{fig:Rheology}a shows the strain amplitude dependence of the first harmonic of $G'$ and $G''$ in oscillatory tests at a frequency \fs{$\omega = 6.28 \rs$}. Data have been normalized \resub{and averaged} over several samples; typical values of the elastic modulus are in the range $20-200$ Pa (see \SI). Both $G'$ and $G''$ are independent of the strain amplitude up to $\gamma \approx 10\%$, beyond which the gel gradually deviates from linear response, with a global tendency towards fluidization. Up to $\gamma = 20\%$, deviations of the viscoelastic moduli with respect to their $\gamma\rightarrow 0$ value are smaller than 10\%, as confirmed by measurements at various $f$, see \SI. We probe the gel response on a wider range of time scales in stress relaxation tests, where a step strain of amplitude $\gamma_0$ is applied at $t=0$ and $\sigma(t)$, the time evolution of the stress needed to maintain such a deformation, is followed for up to 2000 s. Figure \ref{fig:Rheology}b shows the relaxation modulus $G(t) = \sigma(t)/\gamma_0$ for four strain amplitudes $\leq 20\%$. The decay of $G(t)$ is close to logarithmic, confirming a wide distribution of relaxation times, a behavior similar to that reported in other soft solids (alginate gels~\cite{pastore2020anomalous, siviello2015analysis, siviello2016analysis}, granular media under compression~\cite{hartley2003logarithmic, brujic2005granular}, colloidal glasses in creep tests \cite{siebenburger2012creep, suman_analyzing_2019}). While the applied strain changes by a factor of 40, all $G(t)$ curves superimpose, indicating \rrs{linear viscoelastic behavior} up to $\gamma_0 = 20\%$. This is also confirmed by the amplitude of higher-order harmonics in oscillatory tests, a quantity widely used to characterize non-linear behavior ~\cite{hyun2007fourier,hyun2011review}, which shows no significant strain dependence up to $\gamma \gtrsim 30\%$, see \SI. Thus, rheology data collectively indicate marginal, if any, deviations from \rrs{linear viscoelastic} behavior for $\gamma \le 20\%$.

\begin{figure}[t]
    \centering
    \includegraphics[width=.9\columnwidth]{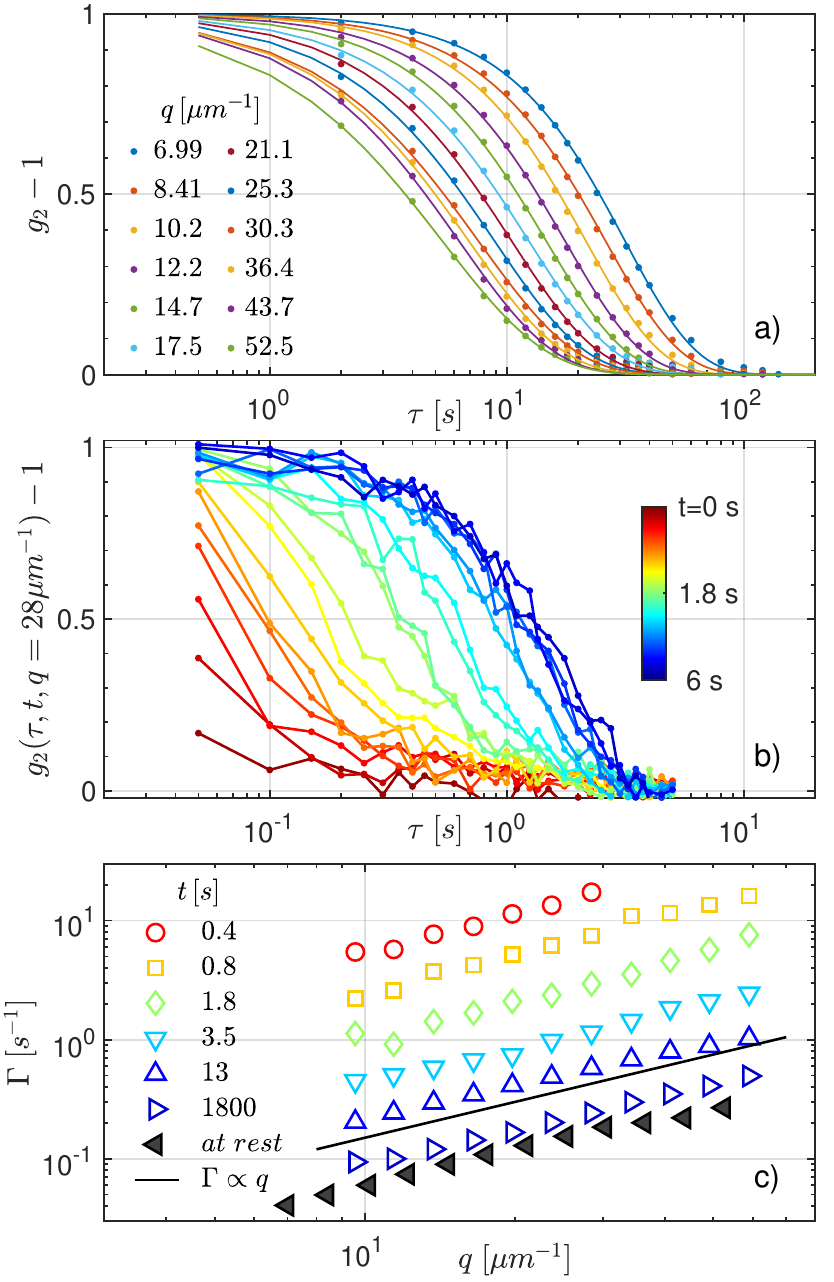}
    \caption{\textbf{Spontaneous and stress-induced dynamics probed by XPCS}. a) Intensity correlation functions for a mucus gel at rest, probed on length scales $q^{-1}$ from $0.053\um$ to $0.33\um$. b) Two-times intensity correlation functions for $q=28.5 \um^{-1}$ display a transient acceleration following a step-strain with $\gamma_0=14.25 \%$. Color code: time $t$ after applying the step strain. c) Open symbols: decay rate of $g_2-1$ \textit{vs} wave-vector $q$, for various $t$ as defined in b). Black solid triangles: relaxation rate for the sample at rest of panel a). The line shows the $\Gamma \propto q$ scaling expected for ballistic dynamics.}
    \label{fig:ESRF}
\end{figure}

\subsection*{Applying a shear strain dramatically accelerates the microscopic dynamics}
Figure~\ref{fig:ESRF}a shows the microscopic dynamics of a mucus gel at rest (no applied strain), as probed by \fs{X-ray photon correlation spectroscopy} (XPCS, see \textit{Material and Methods}). Intensity correlation functions $g_2(\tau)-1$ are measured at several scattering vectors $q$~\cite{Berne_Pecora_1976}. This allows us to probe the relaxation time of the gel density fluctuations on length scales $\sim \pi/q$ spanning almost one decade, from $0.053\um$ to $0.33\um$, smaller than but close to $\approx 3 \um$, \rrs{the length scale beyond which the gel structure changes from fractal-like to rather uniform}~\cite{Philippe2017}.  The full decay of $g_2(\tau)-1$ indicates that the network bonds are not permanent, consistently with the scenario based on previous measurements on mucus gels at lower $q$ vectors~\cite{Philippe2017}. Data are well fitted by a generalized exponential decay, $g_2(\tau)-1 = \exp[- (\Gamma \tau)^\beta \ln 2 ]$, where $\Gamma$ is the half-decay rate defined by $g_2(1/\Gamma)-1 = 0.5$ and $\beta$ controls the shape of the decay. Over the probed $q$ range, we find $\beta  = 1.22 \pm 0.15$ and \fs{$\Gamma \propto q^{a}$, with $a = 0.94 \pm 0.06$} (solid triangles in Fig.~\ref{fig:ESRF}c). Both the compressed exponential shape ($\beta >1$) and the \fs{nearly} linear dependence of the relaxation rate with $q$ \fs{($a \approx 1$)} have been reported for a variety of soft solids, including biological gels~\cite{Cipelletti_Manley_Ball_Weitz_2000, madsen_beyond_2010, Philippe2017, Lieleg_Kayser_Brambilla_Cipelletti_Bausch_2011,bouzid_elastically_2017,pastore2020anomalous}. They are indicative of ballistic dynamics, as opposed to the diffusive motion usually observed \rrs{in polymeric and colloidal systems at thermodynamic equilibrium}~\cite{Berne_Pecora_1976,tang_anomalous_2015,mahmad_rasid_anomalous_2021}, and have been attributed to the slow relaxation of internal stresses in amorphous, out-of-equilibrium soft solids~\cite{Cipelletti_Manley_Ball_Weitz_2000,Bouchaud_Pitard_2001,bouchaud_anomalous_2008}. \rrs{Consistent with this picture, we have shown in previous work that the spontaneous dynamics of mucus slow down over several hours~\cite{Philippe2017}, a behavior known as physical aging and typical of out-of-equilibrium amorphous materials}. On time scales shorter than those accessible to XPCS, thermal fluctuations induce overdamped fluctuations of the gel at fixed network connectivity, in analogy to colloidal and polymeric gels~\cite{Krall1998,Barretta2000,Usuelli2020}. Dynamic light scattering (DLS) reveals that these fast relaxation modes have a characteristic time $ \lesssim 1$ ms and that they account for less than 15\% of the full relaxation of $g_2-1$, see~\SI.

Upon applying a step strain $\gamma_0 = 14.25\%$, within the \rrs{linear viscoelastic} regime, the mucus microscopic dynamics are dramatically enhanced. This is exemplified by Fig.~\ref{fig:ESRF}b, which displays two-times intensity correlation functions (see \textit{Material and Methods}) at a fixed $q$ vector, for various times $t$ after the step strain. We carefully checked that this acceleration does not stem from a spurious motion of the rheometer tool. Immediately after shearing the gel ($t = 0.05$ s, dark red curve in Fig.~\ref{fig:ESRF}b), the dynamics are so fast that the decay of $g_2-1$ is barely measurable; subsequently, the decay rate progressively decreases, approaching that of a gel at rest. The dramatic impact of the applied strain on the microscopic dynamics has to be contrasted with the unchanged mechanical properties of the gel, since $\gamma_0 = 14.25\%$ falls within the linear viscoelastic regime (Fig.~\ref{fig:Rheology}b). Figure~\ref{fig:ESRF}c) shows that the applied strain transiently accelerates the dynamics by more than a factor of 50 at all $q$, i.e. at all probed length scales. Remarkably, the same dependence of the relaxation rate with $q$ is seen during the dynamic acceleration as for the unperturbed gel\fs{, since we find $ \Gamma \propto q^a$ with $a = 1.02 \pm 0.1$ averaged over the datasets with open symbols of Fig.~\ref{fig:ESRF}b}. This suggests that a similar mechanism may be responsible for the dynamics in both cases, i.e. the relaxation of stress acting on the gel, be it internal (as for the unperturbed samples) or externally applied.

\subsection*{Microscopic dynamics correlate with stress relaxation}
To elucidate the relationship between microscopic dynamics and stress relaxation, we perform simultaneous rheology and dynamic light scattering measurements on mucus gels using a custom setup~\cite{Pommella_Philippe_Phou_Ramos_Cipelletti_2019} (see \MM) that probes a scattering vector $q= 33\um^{-1}$, comparable to those in the XPCS experiments. Unlike in microrheology experiments, the DLS measurements probe the \rrs{mucus gels with no added tracer particles. This avoids complications in the data analysis arising when the tracer particles are not fully slaved to the network dynamics, e.g. if they diffuse through the gel pores \cite{weigand_active_2017,wagner_rheological_2017,meldrum_mucin_2018-1}.}

\subsubsection*{Strain ramps: effect of strain rate} 
\begin{figure}[ht!]
    \centering
    \includegraphics[width=\columnwidth]{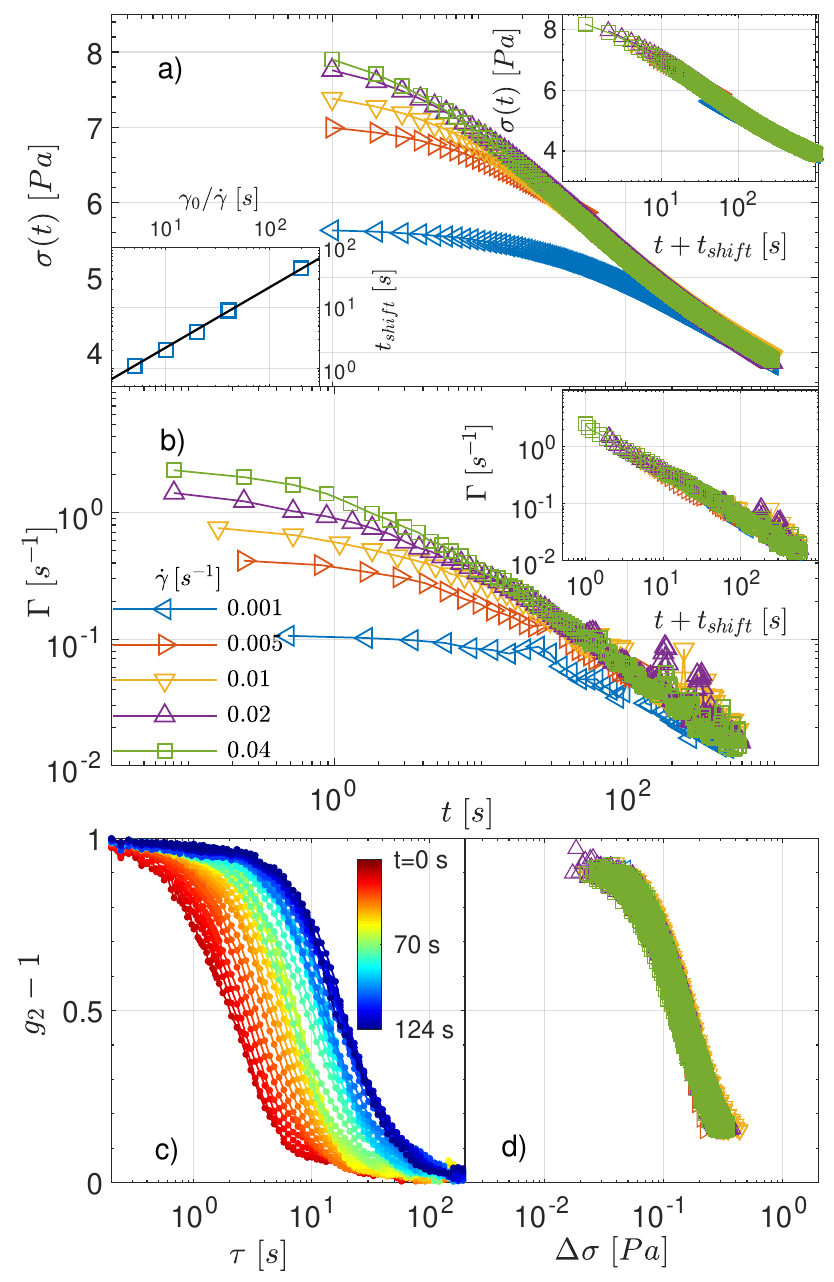}
    \caption{\textbf{Enhanced microscopic dynamics following a short strain ramp at variable strain rate}. Rheology and microscopic dynamics in stress relaxation tests after reaching a fixed strain increment $\gamma_0 = 20 \%$ through strain ramps at various rates $0.001 \le \dot\gamma \le 0.04~\mathrm{s}^{-1}$. In all panels, $t=0$ at the end of the ramp. a) Time-dependent stress relaxation; upper inset: collapse of the same data when plotted \textit{vs} the effective time $t + t_{shift}$. \fs{$t_{shift}$ is proportional to the time $\gamma_0/\dot\gamma$ spent during the ramp, as shown by the lower inset, where the line is a linear fit to the data, $t_{shift} = 0.22\gamma_0/\dot\gamma$}. b) Relaxation rate of the microscopic dynamics measured by DLS at $q = 33 \um^{-1}$ after the strain increment, color code as in a). Inset: same data plotted \textit{vs} shifted time, using the same $t_{shift}$ as in a). c) Representative correlation functions displaying faster decays after a strain ramp \resub{at $\dot\gamma=0.005~\mathrm{s}^{-1}$.} d) \resub{In the accelerated regime ($\Gamma \ge 0.09~\mathrm{s}^{-1}$),} the intensity correlation functions following ramps at all $\dot\gamma$ collapse onto a master curve when plotted \textit{vs} the stress drop $\Delta\sigma = \sigma(t) - \sigma(t+\tau)$. }
    \label{fig:RheoDLSRamp}
\end{figure}

In a first series of experiments, we submit the mucus gels to strain ramps attaining the same final amplitude, $\gamma_0 = 20\%$, but at various  strain rates $0.001~\mathrm{s}^{-1} \le  \dot\gamma \le 0.04~\mathrm{s}^{-1}~$. Figure~\ref{fig:RheoDLSRamp}a shows the stress relaxation following the strain ramp, with $t=0$ the time at which the final strain is attained. At large $t$, all data follow the same trend, close to the logarithmic decay seen in Fig.~\ref{fig:Rheology}b for a step strain. At earlier times, $\sigma$ tends to a plateau, which becomes more pronounced as $\dot\gamma$ decreases. This behavior is rationalized by recalling that bonds within the mucus gel continuously break and reform, and that stress relaxation occurs on a wide range of time scales. Accordingly, part of the stress generated during the ramp is actually relaxed before attaining the final deformation, through the fastest relaxation mechanisms. This scenario is supported by the fact that all the $\sigma(t)$ data collapse onto a master curve when plotting the stress as a function of an effective relaxation time $t + t_{shift}(\dot\gamma)$, with $t_{shift}$ proportional to the time \fs{$\gamma_0/\dot\gamma$} spent during the ramp \fs{(insets of Fig.~\ref{fig:RheoDLSRamp}a)}.

We now turn to the microscopic dynamics. Figure~\ref{fig:RheoDLSRamp}c shows an example of enhanced dynamics, representative of the general behavior. We find the relaxation rate to be markedly accelerated right after attaining the final strain amplitude, after which the microscopic dynamics slow down, in qualitative analogy to the XPCS measurements following a step strain, Fig.~\ref{fig:ESRF}b.  We plot in Fig.~\ref{fig:RheoDLSRamp}b the time-dependent relaxation rate of the microscopic dynamics, for all ramps. Remarkably, $\Gamma(t)$ exhibits the same behavior as the stress relaxation: all data collapse at large $t$, while they tend to plateau at earlier times. As for $\sigma(t)$, the $\Gamma$ plateau is more pronounced for the slower ramps. Finally, the inset of Fig.~\ref{fig:RheoDLSRamp}b shows that all the microscopic dynamics data collapse onto a master curve when plotting $\Gamma$  \textit{vs} the effective relaxation time, using the same time shifts $t_{shift}$ determined for the rheology data.

The strong analogies between the time evolution of the stress and that of the microscopic relaxation rate suggest that the macroscopic mechanical relaxation and the microscopic dynamics are intimately related. We make this observation quantitative by plotting in Fig.~\ref{fig:RheoDLSRamp}d the two-times correlation functions $g_2(t,\tau)-1$ \resub{in the accelerated regime, defined by $\Gamma \ge 0.09~\mathrm{s}^{-1}$,} as a function of the stress drop $\Delta \sigma = \sigma(t) - \sigma(t+\tau)$, rather than the time delay $\tau$.  Remarkably, data for all $\dot\gamma$ and all $t$ during the acceleration phase collapse onto a master curve, demonstrating that the microscopic dynamics depend only on the stress drop, regardless of the strain history imposed to the sample. At longer times, \resub{when $\Gamma$ drops below $0.09~\mathrm{s}^{-1}$, the collapse of $g_2-1$ with $\Delta \sigma$ doesn't hold anymore, suggesting that the externally imposed stress has sufficiently relaxed for the microscopic dynamics to be dominated by the underlying spontaneous dynamics, which is ruled by the relaxation of internal stress, as in gels at rest}.

\subsubsection*{Strain steps: effect of strain amplitude}
We establish the generality of the relationship between stress relaxation and microscopic dynamics in mucus gels by measuring the time evolution of both quantities after imposing a step strain of variable amplitude, $0.24\% \le \gamma_0 \le 9.84\%$, well within the \rrs{linear viscoelastic} regime. The experiments are performed simultaneously on the same sample, taking advantage of the plate-plate geometry of our setup, where the local strain varies linearly with distance from the rotation axis, and where the local dynamics can be measured by space-resolved DLS~\cite{duri_resolving_2009,Pommella_Philippe_Phou_Ramos_Cipelletti_2019} (see \MM).

\begin{figure}[ht]
    \centering
    \includegraphics[width=\columnwidth]{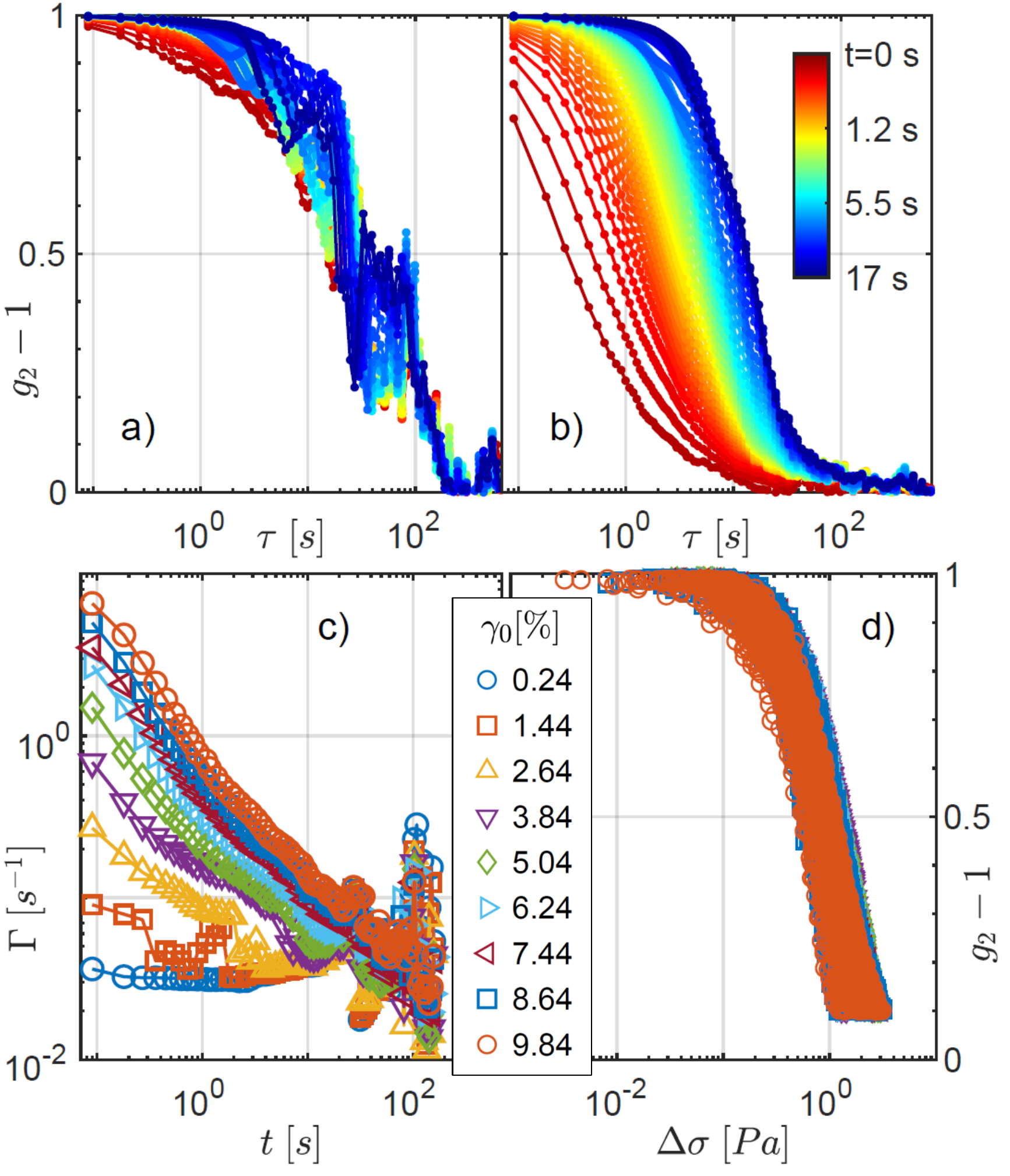}
    \caption{\textbf{Enhanced microscopic dynamics after strain steps of various amplitude.} a),b): Intensity correlation functions at $q = 33 \um^{-1}$ after strain steps with $\gamma_0 = 0.24\%$ (a) and $\gamma_0 = 7.44\%$ (b). The time $t$ after the step is indicated in both panels by the color code of the bar in b). c): Time dependence of the relaxation rate of the microscopic dynamics, after strain steps at various $\gamma_0$, \resub{shown in \% by the labels between panels c) and d)}. d): Master curve when plotting all correlation functions in the accelerated regime \resub{($\Gamma \ge 0.09~\mathrm{s}^{-1}$ and $\gamma_0\ge 1.44\%$)} \textit{vs} the stress drop $\Delta\sigma = \sigma(t) - \sigma(t+\tau)$. \resub{Same symbols as in c)}.}
    \label{fig:RheoDLSStep}
\end{figure}

Figure~\ref{fig:RheoDLSStep} contrasts the dynamics for the smallest strain, $\gamma_0 = 0.24\%$ (panel a), with those for one of the largest strains, $\gamma_0 = 7.44\%$ (panel b). For the former, the applied step strain has no measurable effect on the decay of $g_2-1$, which occurs at the same rate as for the unperturbed sample and exhibits strong fluctuations, a distinctive feature of the spontaneous dynamics of mucus gels~\cite{Philippe2017}. At larger $\gamma_0$, by contrast, the dynamics are transiently enhanced, similarly to the XPCS and strain ramp experiments, Figs.~\ref{fig:ESRF} and ~\ref{fig:RheoDLSRamp} respectively. The temporal evolution of the microscopic relaxation rate $\Gamma$ is shown in Fig.~\ref{fig:RheoDLSStep}c for all the tested $\gamma_0$. The transient acceleration lasts about 15 s; its magnitude grows with the amplitude of the applied strain, up to more than a 100-fold increase of the microscopic relaxation rate for $\gamma_0 = 9.84\%$ and $t=0.1$ s. Figure ~\ref{fig:RheoDLSStep}d shows that all correlation functions measured during the accelerated phase collapse onto a single master curve when plotting them \textit{vs} the stress decay $\Delta \sigma$, regardless of the applied strain. Thus, the scaling of $g_2-1$ with $\Delta \sigma$ is robust not only with respect to a change of the strain ramp rate, as demonstrated by Fig.~\ref{fig:RheoDLSRamp}, but also upon changes of the amplitude of the imposed strain.

\section*{Modelling the relationship between microscopic dynamics and stress relaxation}

We propose a simple model that rationalizes the relationship between the microscopic dynamics and the macroscopic stress relaxation. The main ingredients of the model are outlined here: see \MM~and \SI~for more details. \resub{The model focuses on the accelerated network dynamics upon applying a shear strain; it neglects the fast dynamics ($\Gamma \gtrsim 100~\mathrm{s}^{-1}$) due to \rrs{gel} fluctuations at fixed network connectivity, as well as the spontaneous slow dynamics ($\Gamma \lesssim 0.09~\mathrm{s}^{-1}$) that occur even in the absence of an applied strain}. As indicated by the spontaneous dynamics \rrs{and the rheology data, bonds within the gel network} are continuously broken and reformed. In between a bond breaking and the following bond formation, the network relaxes its configuration so as to minimize the elastic energy. We shall term “rearrangement event” the sequence bond breaking, network relaxation, \rrs{bond formation. Note that bonds will generally be reformed in a different microscopic configuration, as indicated by the decay of both the intensity correlation function and the macroscopic stress}. Under an applied strain, after $n$ events the macroscopic stress measured by the rheometer drops by an amount $\Delta \sigma = n \delta\bar\sigma$, with $\delta \bar\sigma$ the average drop per event of the macroscopic stress. We further assume $\delta \bar\sigma$ to be proportional to the macroscopic stress acting on the gel: $\delta \bar\sigma = B \sigma(t)$, with $B <<1$ a numerical prefactor. $B$ accounts for the fraction of the sample volume that no more contributes to the elastic response of the gel after one rearrangement event. This expression may be further simplified to $\delta \bar\sigma = B \gamma_0 G_0$, since we focus on the accelerated regime, at small $t$, over which the macroscopic stress drop is small compared to the initial stress (typically, a fraction of Pa \textit{vs} several Pa, see Figs.~\ref{fig:RheoDLSRamp}a,d and~\ref{fig:RheoDLSStep}d), such that $\sigma(t) \approx \sigma(0) = \gamma_0 G_0$, with $G_0$ the gel elastic modulus.

Each event entails the elastic relaxation of the network, whose \rrs{components} undergo a microscopic root mean square displacement $\delta\bar r$. After $n$ events, the cumulated root mean square displacement is $\Delta r = n^p \delta\bar r$, where the exponent $p$ accounts for the nature of the dynamics. Two limiting cases are diffusive dynamics, where the displacements due to successive events are totally uncorrelated ($p=0.5$), and ballistic dynamics, where subsequent events locally displace the network along the same direction ($p=1$). Finally, we assume that the microscopic response to a rearrangement event is ruled by linear elasticity, implying $\delta\bar r = \frac{L}{G_0} \delta\bar\sigma$, where $L$ is a microscopic length scale.

The decay of the correlation function after $n$ events leading to a typical displacement $\Delta r$ may quite generally be written as $g_2(\Delta r)-1 = \exp \left [ -A(q n^p \delta\bar r )^{\beta} \right ]$, where $A$ and $\beta$ are parameters of order unity accounting for the probability distribution of the displacements (see~\SI). Using the above expressions, we recast $g_2-1$ as a function of the macroscopic stress drop and imposed strain:
\begin{equation}
    g_2(\Delta\sigma,\gamma_0)-1 = \exp \left[ - A(Lq)^\beta \left ( \frac{\Delta \sigma}{G_0} \right )^{p \beta} (\gamma_0 B)^{\beta(1-p)} \right] \,.
    \label{eq:model-general}
\end{equation}
This general result considerably simplifies for ballistic dynamics ($p=1$), since the explicit dependence on $\gamma_0$ drops out and Eq.~\ref{eq:model-general} reduces to
\begin{equation}
    g_2(\Delta\sigma)-1 = \exp \left[ - \left ( \frac{qL'}{G_0} \right )^{\beta}  \Delta\sigma^{\beta} \right] \,,
    \label{eq:g2_Dsigma}
\end{equation}
where we have incorporated the constant $A$ into the characteristic length $L'\equiv A^{1/\beta} L$. In our experiments, we find that $g_2-1$ only depends on $\Delta \sigma$, not on $\gamma_0$, see Fig.~\ref{fig:RheoDLSStep}. Furthermore, in the XPCS experiments we find $\Gamma \sim q$. Both results are consistent with ballistic dynamics. In the following, we shall thus take $p=1$ and use the simpler form of the model, Eq.~\ref{eq:g2_Dsigma}.

\subsection*{Spring network simulations validate the model assumptions}
\begin{figure*}[h!]
    \centering
    \includegraphics[width=\linewidth]{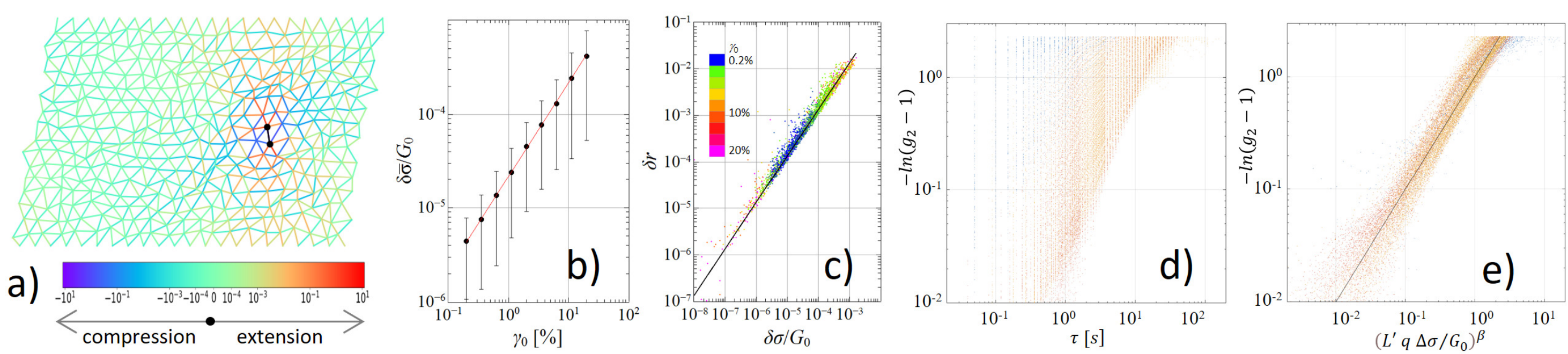}
    \caption{\textbf{Spring network simulations and comparison with the experiments validate the model}. a): Snapshot of a network of springs subject to a macroscopic strain $\gamma_0=20\%$. The springs are shown as lines whose color indicates the local strain (in \%) following the relaxation of the spring in black, to mimic a bond breaking and reforming event. b) Linear dependence of the normalized stress drop per event on the applied strain. c) Linear dependence of the rms displacement of network nodes \textit{vs} normalized stress drop. d) Experimental intensity correlation functions \textit{vs} time delay $\tau$ during the enhanced dynamics phase, in a representation convenient for the comparison with the model. Blue, red and yellow shades are XPCS, DLS strain ramp, and DLS step strain data, respectively. e) Same data as in d), plotted as a function of the normalized stress drop, using the $L'$ and $\beta$ values shown in Table~\ref{tab:fitparameters}. All data collapse on the line that shows the model prediction.}
    \label{fig:model}
\end{figure*}

To test the key assumptions used to derive the model, \resub{$\delta \bar r \propto \delta \bar \sigma/G_0$ and $\delta \bar \sigma \propto \gamma_0 G_0$}, we simulate the gel elastic response to a \resub{single} rearrangement event using 2D spring networks, Fig.~\ref{fig:model}a. In the simulations, an event consists in passivating a randomly chosen spring, relaxing the network, and restoring the spring in a neutral configuration (see \MM~for details). Figure~\ref{fig:model}b demonstrates that $\delta \bar\sigma = B \gamma_0 G_0$, as assumed in the model. Furthermore, we find $B = 2.15\times 10^{-3}$ (expressing strain in absolute units), of the same order of magnitude as $0.91\times 10^{-3}$, the fraction of the sample volume that is relaxed by passivating one out of the 1100 simulated springs. Figure~\ref{fig:model}c shows the linear relationship between the microscopic displacement upon a rearrangement event and the associated macroscopic stress drop, by displaying the individual displacements $\delta r$ for 2668 rearrangement events in networks under strains $0.2 \% \le \gamma_0 \le 20\%$. 
By fitting the data to a straight line, $\delta r = L \delta \sigma/G_0$, we find $L = 13.05$, in units of the network mesh size. Finally, the simulations show that $\delta r$ is isotropic and not preferentially oriented along the shear direction, see \SI.

\subsection*{Comparison with experiments}
We  fit Eq.~\ref{eq:g2_Dsigma} to the data, with $\beta$ and $L'$ the fitting parameters and $G_0$ as directly measured by rheology. The same $\beta$ and $L'$ parameters are shared among all data collected in each of Figs.~\ref{fig:ESRF},~\ref{fig:RheoDLSRamp} and ~\ref{fig:RheoDLSStep}, respectively. Figure~\ref{fig:model}d shows the raw correlation functions, in a representation convenient for the comparison with the model. When plotted against the time delay $\tau$, the correlation functions are spread over more than two decades in relaxation rate. Figure~\ref{fig:model}e shows a remarkable collapse of the same data when plotted against $(qL'\Delta \sigma / G_0)^{\beta}$, as implied by Eq.~\ref{eq:g2_Dsigma}. Correlation functions measured over a factor of 6 in $q$ vectors, a factor of 40 in $\dot\gamma$ and two decades in $\gamma_0$ all agree with the model, shown by the straight line in Fig.~\ref{fig:model}e. 
Table~\ref{tab:fitparameters} shows the fitting parameters $\beta$ and $L'$: similar  values are found across all experiments, lending further support to the model.

\begin{table}
\centering
\caption{\textbf{\fs{Model parameters$^a$}} for the experiments of Figs.~\ref{fig:ESRF}--\ref{fig:RheoDLSStep}}
\begin{tabular}{lrrr}
Parameter\fs{$^a$} & XPCS, Fig.~\ref{fig:ESRF} & DLS, Fig.~\ref{fig:RheoDLSRamp} & DLS, Fig.~\ref{fig:RheoDLSStep}\\
\midrule
$\beta$ & \resub{$1.46\pm0.1$} & \resub{$1.51\pm0.1$} & \resub{$1.33\pm0.1$} \\
$L' [\mu m]$ & \resub{$12.3\pm 1$} & \resub{$5.6\pm0.6$} & \resub{$4.3\pm 0.4$} \\
$G_0~\mathrm{[Pa]}$ & $53.6$ & $45$ & $180$ \\
\bottomrule
\end{tabular}
\newline
\addtabletext{\fs{$^a$$\beta$ and $L'$ are fitting parameters, $G_0$ is a fixed parameter obtained directly from rheology.}}
\label{tab:fitparameters}
\end{table}

\section*{Discussion and Conclusions}
The central result of our study is the dramatic enhancement of the microscopic mucus dynamics upon applying a modest shear stress. The mucus mobility increases by up to two orders of magnitude, and slowly relax to its unperturbed value over several tens of seconds. Bonds in mucus gels continuously break and reform. At first sight, the most natural explanation for the accelerated dynamics would be stress-enhanced bond dynamics, a key feature of non-covalent biomolecular interactions~\cite{evans_introductory_1998,Lieleg_Kayser_Brambilla_Cipelletti_Bausch_2011}. More generally, amorphous soft solids exhibit accelerated microscopic dynamics upon applying a strong mechanical drive~\cite{viasnoff_rejuvenation_2002,schall_structural_2007,lee_direct_2009,warren_deformation-induced_2010-1,rogers_echoes_2014,sentjabrskaja_creep_2015,2020SciPy-NMeth,aime_microscopic_2018,rogers_microscopic_2018}, a phenomenon known as `rejuvenation' and captured by models such as the soft glassy rheology~\cite{sollich_rheology_1997}. However, stress-induced  bond dynamics and rejuvenation impact both the microscopic dynamics and the macroscopic mechanical response: as such, they are distinctive features of the non-linear viscoelastic regime. By contrast, the enhanced dynamics reported here occur in the \rrs{linear viscoelastic} regime. Our model rationalizes them with no need of invoking enhanced bond dynamics, which would be incompatible with linear viscoelasticity. \resub{The key ingredients of the model are network elasticity and bond breaking: the dynamics are due to the elastic strain field generated within the network when a bond is broken.} Our experiments indicate that the dynamics are ballistic. This suggests correlations between successive rearrangement events, resulting in \rrs{local displacements of the gel network} along the same direction over several events, in analogy to \rrs{modelling and numerical results} for the spontaneous dynamics of gel networks due to internal stresses~\cite{Duri_Cipelletti_2006,bouzid_elastically_2017}.

By fitting the model to the data, we identify a length scale $L$ of the order of several microns that characterizes the elastic propagation of the microscopic strain field set by one single event. Interestingly, this length scale is close to important structural length scales highlighted in previous works. In Ref.~\cite{Philippe2017}, pig gastric mucus was shown to exhibit fractal morphology up to $\approx 3 \um$, beyond which its structure became rather uniform. The largest pores in respiratory mucus have a comparable size~\cite{kirch_optical_2012-2}. Pig mucus covering the airways is mainly composed by gel-forming
mucins MUC5B and MUC5AC, which were recently shown to be produced by goblet cells and submucosal glands in distinct morphological structures (strands, threads, and sheets), with cross section on the order of a few microns~\cite{ostedgaard_gel-forming_2017}. Our results suggests that these structural length scales may also have a relevance for the mechanical response of mucus at the microscopic level. Note that our model invokes linear elastic response: the emerging length scale $L$ is also consistent with \rrs{$\approx 10\um$}, the crossover length scale above which microrheology experiments recover the macroscopic mechanical behavior~\cite{lai_micro-_2009,weigand_active_2017,meldrum_mucin_2018-1}.

Our finding that even a modest stress greatly enhances the microscopic dynamics may help understanding why on micron scales mucus is reorganized to a greater extent compared to expectations from macroscopic rheology. Celli \textit{et al.} estimate the stress exerted by the mucus-penetrating bacterium \textit{Helicobacter Pylori} to be around 1 Pa, one decade smaller than the macroscopic yield stress of mucus~\cite{celli_helicobacter_2009}. The stress-induced enhancement of the mucus dynamics reported here may be an additional factor allowing for bacterial penetration, together with the alteration of the rheological properties of mucus due to chemicals released by the bacterium~\cite{celli_helicobacter_2009}. Other processes for which a stress-induced enhancement of the microscopic dynamics may be relevant include cilia beating and the transport through mucus of nanoparticle-based drug vectors. The mechanisms invoked to rationalize our findings rely only on the notion of non-permanent bonds and linear elasticity: we thus expect the phenomenology reported here to be generic to \rrs{out-of-equilibrium} physical gels. More generally, we expect similar mechanisms to be relevant to the microscopic dynamics of mechanically driven soft solids, in both the linear and non-linear \rrs{viscoelastic} regime, since any local rearrangement will entail an elastic strain field whose magnitude \fs{depends} on the externally applied stress. Further experiments will be needed to test these hypotheses in mucus gels and other soft solids.








\matmethods{\subsection*{Mucus samples}
Samples were collected from the stomach of just-slaughtered pigs, extensively washed with water and immediately frozen, with no further purification or homogenization treatments. Before testing, a fragment of the stomach was thawed and scraped to collect the sample. Sodium azide ($0.02\%~\mathrm{w/w}$) was added to prevent bacterial growth. \resub{The sample pH was $5.8 \pm 0.5$ and the dry fraction ranged from 9\% to 15.5\%, depending on sample.}

\subsection*{Rheology}
The measurements of Fig.~\ref{fig:Rheology} were performed on a Thermo Scientific Haake Mars III rheometer, equipped with a cone and plate tool (cone angle: $1^{\circ}$, diameter 30 mm). Stress relaxation data simultaneous to the XPCS measurements of Fig.~\ref{fig:ESRF} were obtained using a Haake RS6000 rheometer, equipped with a plate-plate tool (plate diameter: $2R = 20$ mm, gap: $e = 1$ mm \cite{zinn2018}). All other data were obtained on an Anton Paar MCR502 rheometer equipped with glass plate-plate tools, with $2R = 50$ mm and $e = 0.3$ mm. In all tests, mucus was loaded in the rheometer preheated at $37^{\circ}~\mathrm{C}$ and left at rest at least 20 minutes to erase any stress induced by loading. A thin layer of silicone oil was deposited on the sample rim, to prevent water evaporation. Note that in the plate-plate geometry, both the strain and the stress increase linearly with distance from the tool axis \cite{macosko1994rheology}. In this work, $\gamma_0$ and $\sigma$ refer to the strain and stress at the location where the microscopic dynamics are measured.

\subsection*{Rheo-XPCS and rheo-DLS measurements}
XPCS measurements were performed at the ID02 beamline of the European Synchrotron Radiation Facility (ESRF), using a partially coherent X-ray beam with wavelength $\lambda = 9.95 \times 10^{-2}$ nm and cross-section $20 \times 20\um^2$, and a Eiger 500K camera at distance $d$ from the sample as a detector. The mucus spontaneous dynamics (Fig.~\ref{fig:ESRF}a) were measured at $d = 20~\mathrm{m}$, by loading the sample in a glass capillary. Coupled rheo-XPCS measurements were performed in the tangential geometry, where the X-ray beam passes through the sample near the edge of the plate-plate tool (see \SI). For the rheo-XPCS measurements, $d$ was set to $30.7~\mathrm{m}$ and a small amount of silica particles was added to the sample to enhance the scattering contrast (Ludox TM50 by Aldrich, radius $\sim 18~\mathrm{nm}$ \cite{Truzzolillo2015}, 1.25\% w/w), with no change of the rheological properties (\SI). Images of the scattered light were processed according to standard methods (see Ref.~\cite{Duri_Bissig_Trappe_Cipelletti_2005} and \SI) to calculate the two-time intensity correlation $g_2(t,\tau,q)-1$, averaged over a set of scattering vectors $\mathbf{q}$ with nearly the same magnitude $4\pi \lambda^{-1} \sin(\theta/2)$ but different azimuthal orientation, with $\theta$ the scattering angle.

Rheo-DLS measurements were performed on bare mucus (no added particles), using a custom setup~\cite{Pommella_Philippe_Phou_Ramos_Cipelletti_2019}, as detailed in the \SI. In brief, the sample was illuminated through the transparent bottom plate of the rheometer by a laser beam with $\lambda = 532.5~\mathrm{nm}$. Images of the backscattered light were collected by a CMOS camera, corresponding to a scattering vector $\mathbf{q}$ whose vertical component, parallel to the rheometer axis, accounts for 90\% of the overall magnitude of $\mathbf{q}$, $q = 33\um^{-1}$.
For the data of Fig.~\ref{fig:RheoDLSRamp}, two-time intensity correlation functions $g_2(t,\tau,q=33\um^{-1})$ were averaged over the full field of view, corresponding to a cylindrical sample volume of radius 5 mm and thickness $e=0.3$ mm, located at $\simeq17$ mm from the tool axis. The data of Fig.~\ref{fig:RheoDLSStep} were collected by imaging the whole rheometer plate. The correlation functions where averaged over nine rings of pixel corresponding to growing values of the radial distance from the rheometer axis and thus of $\sigma$ and $\gamma_0$.

\subsection*{Simulations}

Two-dimensional spring networks are used to simulate the response of a sheared elastic network to a bond breaking event. The networks comprise 1100 springs, initially placed on a triangular lattice with lattice parameter $a$. Periodic boundary conditions are implemented in the $x$ direction. The nodes of the first and last row have fixed $y$ coordinates, $y=0$ and $y= y_{max}$, respectively, and are connected only to nodes in the bulk. Initial configurations are obtained by displacing in the $x$ direction the upper row of nodes by an amount $\gamma_0 y_{max}$, thereby applying a constant shear $\gamma_0$. The network is then relaxed by minimizing the total elastic energy with respect to the position of the nodes in the bulk. The energy minimization is implemented in a custom Python code using the Scipy \textit{minimize} function~\cite{2020SciPy-NMeth}. Disorder is introduced by randomly drawing the spring rest lengths $l_0$ and spring constants $k$ from Gaussian distributions, with $<l_0> = a$ and typical relative standard deviation $\sigma_{l_0}/a = \sigma_{k}/<k>=0.25$. We check that the results do not depend on the details of the PDF, in the limit of moderate disorder (relative standard deviation $< 0.5$).

Rearrangement events are simulated as pairs of bond breaking and \rrs{bond forming} events. Bond breaking is mimicked by randomly choosing a spring $i$ of the sheared system, setting $k_i=0$ and relaxing the network configuration. The same bond is then reformed in an unstrained state, by setting the spring rest length to the bond length in the new configuration. Finally, the spring constant $k_i$ is set back to its value prior to the rearrangement event. This procedure mimics \rrs{new bonds that, following a bond breaking event, form in a relaxed state and thus do not require elastic energy to be input to the system. These new bonds, however,} will contribute to the network elasticity in response to further bond breaking events. We measure the change of the shear stress, $\delta \sigma_{xy}$, and the rms displacement of the bulk nodes, $\delta r$, following one event. \resub{Here, $\delta r = \sqrt{\delta x^2+\delta y^2}$, with $\delta x$, $\delta y$ the rms displacements of the bulk nodes along the $x$ and $y$ direction, respectively}. The shear stress is defined as the $x$ component of the force per node acting on the upper and lower lattice rows, measured in units of $<k>$. We find that relaxing a spring oriented nearly parallel to the plates entails a negligible stress drop $\delta \sigma_{xy}$. We thus restrict our analysis to rearrangement events involving springs that form an angle larger than $20^{\circ}$ with the plates. We find isotropic response, $\delta x^2 \approx \delta y ^2 \approx 0.5 \delta r^2$.  All displacements are expressed in units of $a$. The results shown in Fig.\ref{fig:model} have been obtained for both pristine networks and networks that previously underwent up to 500 rearrangement events.

}

\showmatmethods{} 

\acknow{\rrs{We thank J.-L. Barrat for insightful discussions}. We acknowledge the European Synchrotron Radiation Facility for provision of synchrotron radiation facilities and we would like to thank T. Narayan and L. Sharpnack for assistance in using beamline ID02. We thank the French CNES, CNRS, ANR (grants No. ANR-14-CE32-0005, FAPRES, and ANR-20-CE06-0028, MultiNet), and ANRT (grant No. 2014/0109) for financial support. LC acknowledges support from the Institut Universitaire de France.}

\showacknow{} 

\bibliography{mucus}

\begin{thebibliography}{10}

\bibitem{denny_invertebrate_1989}
MW Denny, Invertebrate mucous secretions: Functional alternatives to vertebrate
  paradigms.
\newblock {\em\protect\JournalTitle{Symp Soc Exp Biol}} \textbf{43}, 337--366
  (1989).

\bibitem{vasquez2015complex}
PA Vasquez, MG Forest, Complex fluids and soft structures in the human body in
  {\em Complex Fluids in Biological Systems}.
\newblock (Springer), pp. 53--110 (2015).

\bibitem{cone_barrier_2009}
RA Cone, Barrier properties of mucus.
\newblock {\em\protect\JournalTitle{Advanced Drug Delivery Reviews}}
  \textbf{61}, 75--85 (2009).

\bibitem{bansil_mucin_2006}
R Bansil, BS Turner, Mucin structure, aggregation, physiological functions and
  biomedical applications.
\newblock {\em\protect\JournalTitle{Current Opinion in Colloid \& Interface
  Science}} \textbf{11}, 164--170 (2006).

\bibitem{wagner_mucins_2018}
C Wagner, K Wheeler, K Ribbeck, Mucins and {{Their Role}} in {{Shaping}} the
  {{Functions}} of {{Mucus Barriers}}.
\newblock {\em\protect\JournalTitle{Annu. Rev. Cell Dev. Biol.}} \textbf{34},
  189--215 (2018).

\bibitem{ridley2018mucins}
C Ridley, DJ Thornton, Mucins: the frontline defence of the lung.
\newblock {\em\protect\JournalTitle{Biochemical Society Transactions}}
  \textbf{46}, 1099--1106 (2018).

\bibitem{meldrum_mucin_2018-1}
OW Meldrum, et~al., Mucin gel assembly is controlled by a collective action of
  non-mucin proteins, disulfide bridges, {{Ca2}}+-mediated links, and hydrogen
  bonding.
\newblock {\em\protect\JournalTitle{Sci Rep}} \textbf{8}, 5802 (2018).

\bibitem{lai_micro-_2009}
SK Lai, YY Wang, D Wirtz, J Hanes, Micro- and macrorheology of mucus.
\newblock {\em\protect\JournalTitle{Advanced Drug Delivery Reviews}}
  \textbf{61}, 86--100 (2009).

\bibitem{cohn2006mucus}
L Cohn, , et~al., Mucus in chronic airway diseases: sorting out the sticky
  details.
\newblock {\em\protect\JournalTitle{The Journal of clinical investigation}}
  \textbf{116}, 306--308 (2006).

\bibitem{georgiades_particle_2014}
P Georgiades, PDA Pudney, DJ Thornton, TA Waigh, Particle tracking
  microrheology of purified gastrointestinal mucins.
\newblock {\em\protect\JournalTitle{Biopolymers}} \textbf{101}, 366--377
  (2014).

\bibitem{stokes2008rheology}
JR Stokes, WJ Frith, Rheology of gelling and yielding soft matter systems.
\newblock {\em\protect\JournalTitle{Soft Matter}} \textbf{4}, 1133--1140
  (2008).

\bibitem{denny1980physical}
MW Denny, JM Gosline, The physical properties of the pedal mucus of the
  terrestrial slug, ariolimax columbianus.
\newblock {\em\protect\JournalTitle{Journal of experimental Biology}}
  \textbf{88}, 375--394 (1980).

\bibitem{denny1983molecular}
MW Denny, Molecular biomechanics of molluscan mucous secretions in {\em
  Metabolic Biochemistry and Molecular Biomechanics}.
\newblock (Elsevier), pp. 431--465 (1983).

\bibitem{Philippe2017}
AM Philippe, L Cipelletti, D Larobina, Mucus as an arrested phase separation
  gel.
\newblock {\em\protect\JournalTitle{Macromolecules}} \textbf{50}, 8221--8230
  (2017).

\bibitem{macierzanka2014transport}
A Macierzanka, et~al., Transport of particles in intestinal mucus under
  simulated infant and adult physiological conditions: impact of mucus
  structure and extracellular dna.
\newblock {\em\protect\JournalTitle{PloS one}} \textbf{9}, e95274 (2014).

\bibitem{weigand_active_2017}
WJ Weigand, et~al., Active microrheology determines scale-dependent material
  properties of {{Chaetopterus}} mucus.
\newblock {\em\protect\JournalTitle{PLoS One}} \textbf{12}, e0176732 (2017).

\bibitem{wagner_rheological_2017}
CE Wagner, BS Turner, M Rubinstein, GH McKinley, K Ribbeck, A {{Rheological
  Study}} of the {{Association}} and {{Dynamics}} of {{MUC5AC Gels}}.
\newblock {\em\protect\JournalTitle{Biomacromolecules}} \textbf{18}, 3654--3664
  (2017).

\bibitem{demouveaux_gel-forming_2018}
B Demouveaux, V Gouyer, F Gottrand, T Narita, JL Desseyn, Gel-forming mucin
  interactome drives mucus viscoelasticity.
\newblock {\em\protect\JournalTitle{Advances in Colloid and Interface Science}}
  \textbf{252}, 69--82 (2018).

\bibitem{jory_mucus_2019}
M Jory, et~al., Mucus {{Microrheology Measured}} on {{Human Bronchial
  Epithelium Culture}}.
\newblock {\em\protect\JournalTitle{Front. Phys.}} \textbf{7}, 19 (2019).

\bibitem{mason_optical_1995}
TG Mason, DA Weitz, Optical {{Measurements}} of {{Frequency}}-{{Dependent
  Linear Viscoelastic Moduli}} of {{Complex Fluids}}.
\newblock {\em\protect\JournalTitle{Phys. Rev. Lett.}} \textbf{74}, 1250--1253
  (1995).

\bibitem{bajka2015influence}
BH Bajka, NM Rigby, KL Cross, A Macierzanka, AR Mackie, The influence of small
  intestinal mucus structure on particle transport ex vivo.
\newblock {\em\protect\JournalTitle{Colloids and Surfaces B: Biointerfaces}}
  \textbf{135}, 73--80 (2015).

\bibitem{norton2011model}
MM Norton, RJ Robinson, SJ Weinstein, Model of ciliary clearance and the role
  of mucus rheology.
\newblock {\em\protect\JournalTitle{Physical Review E}} \textbf{83}, 011921
  (2011).

\bibitem{figueroa2019mechanical}
N Figueroa-Morales, L Dominguez-Rubio, TL Ott, IS Aranson, Mechanical shear
  controls bacterial penetration in mucus.
\newblock {\em\protect\JournalTitle{Scientific reports}} \textbf{9}, 1--10
  (2019).

\bibitem{celli_helicobacter_2009}
JP Celli, et~al., Helicobacter pylori moves through mucus by reducing mucin
  viscoelasticity.
\newblock {\em\protect\JournalTitle{Proceedings of the National Academy of
  Sciences}} \textbf{106}, 14321--14326 (2009).

\bibitem{king1987role}
M King, The role of mucus viscoelasticity in cough clearance.
\newblock {\em\protect\JournalTitle{Biorheology}} \textbf{24}, 589--597 (1987).

\bibitem{datta_polymers_2016}
SS Datta, A Preska~Steinberg, RF Ismagilov, Polymers in the gut compress the
  colonic mucus hydrogel.
\newblock {\em\protect\JournalTitle{Proc Natl Acad Sci USA}} \textbf{113},
  7041--7046 (2016).

\bibitem{button_periciliary_2012-1}
B Button, et~al., A {{Periciliary Brush Promotes}} the {{Lung Health}} by
  {{Separating}} the {{Mucus Layer}} from {{Airway Epithelia}}.
\newblock {\em\protect\JournalTitle{Science}} \textbf{337}, 937--941 (2012).

\bibitem{anderson_relationship_2015}
WH Anderson, et~al., The {{Relationship}} of {{Mucus Concentration}}
  ({{Hydration}}) to {{Mucus Osmotic Pressure}} and {{Transport}} in {{Chronic
  Bronchitis}}.
\newblock {\em\protect\JournalTitle{Am J Respir Crit Care Med}} \textbf{192},
  182--190 (2015).

\bibitem{ewoldt_rheological_2007}
RH Ewoldt, C Clasen, AE Hosoi, GH McKinley, Rheological fingerprinting of
  gastropod pedal mucus and synthetic complex fluids for biomimicking adhesive
  locomotion.
\newblock {\em\protect\JournalTitle{Soft Matter}} \textbf{3}, 634--643 (2007).

\bibitem{hyun2011review}
K Hyun, et~al., A review of nonlinear oscillatory shear tests: Analysis and
  application of large amplitude oscillatory shear (laos).
\newblock {\em\protect\JournalTitle{Progress in Polymer Science}} \textbf{36},
  1697--1753 (2011).

\bibitem{coblas2016correlation}
D Coblas, D Broboana, C Balan, Correlation between large amplitude oscillatory
  shear (laos) and steady shear of soft solids at the onset of the fluid
  rheological behavior.
\newblock {\em\protect\JournalTitle{Polymer}} \textbf{104}, 215--226 (2016).

\bibitem{bonn_yield_2017}
D Bonn, MM Denn, L Berthier, T Divoux, S Manneville, Yield stress materials in
  soft condensed matter.
\newblock {\em\protect\JournalTitle{Reviews of Modern Physics}} \textbf{89},
  035005 (2017).

\bibitem{donley_elucidating_2020}
GJ Donley, PK Singh, A Shetty, SA Rogers, Elucidating the {$G^{\prime\prime}$}
  overshoot in soft materials with a yield transition via a time-resolved
  experimental strain decomposition.
\newblock {\em\protect\JournalTitle{Proc Natl Acad Sci USA}} \textbf{117},
  21945--21952 (2020).

\bibitem{viasnoff_rejuvenation_2002}
V Viasnoff, F Lequeux, Rejuvenation and overaging in a colloidal glass under
  shear.
\newblock {\em\protect\JournalTitle{Phys. Rev. Lett.}} \textbf{89}, 065701
  (2002).

\bibitem{schall_structural_2007}
P Schall, DA Weitz, F Spaepen, Structural {{Rearrangements That Govern Flow}}
  in {{Colloidal Glasses}}.
\newblock {\em\protect\JournalTitle{Science}} \textbf{318}, 1895--1899 (2007).

\bibitem{rogers_echoes_2014}
MC Rogers, et~al., Echoes in x-ray speckles track nanometer-scale plastic
  events in colloidal gels under shear.
\newblock {\em\protect\JournalTitle{Physical Review E}} \textbf{90}, 062310
  (2014).

\bibitem{sentjabrskaja_creep_2015}
T Sentjabrskaja, et~al., Creep and flow of glasses: Strain response linked to
  the spatial distribution of dynamical heterogeneities.
\newblock {\em\protect\JournalTitle{Scientific Reports}} \textbf{5}, 11884
  (2015).

\bibitem{2020SciPy-NMeth}
P Virtanen, et~al., {{SciPy} 1.0: Fundamental Algorithms for Scientific
  Computing in Python}.
\newblock {\em\protect\JournalTitle{Nature Methods}} \textbf{17}, 261--272
  (2020).

\bibitem{aime_microscopic_2018}
S Aime, L Ramos, L Cipelletti, Microscopic dynamics and failure precursors of a
  gel under mechanical load.
\newblock {\em\protect\JournalTitle{PNAS}} \textbf{115}, 3587--3592 (2018).

\bibitem{rogers_microscopic_2018}
MC Rogers, et~al., Microscopic signatures of yielding in concentrated
  nanoemulsions under large-amplitude oscillatory shear.
\newblock {\em\protect\JournalTitle{Physical Review Materials}} \textbf{2},
  095601 (2018).

\bibitem{pastore2020anomalous}
R Pastore, C Siviello, F Greco, D Larobina, Anomalous aging and stress
  relaxation in macromolecular physical gels: The case of strontium alginate.
\newblock {\em\protect\JournalTitle{Macromolecules}} \textbf{53}, 649--657
  (2020).

\bibitem{siviello2015analysis}
C Siviello, F Greco, D Larobina, Analysis of linear viscoelastic behaviour of
  alginate gels: effects of inner relaxation, water diffusion, and syneresis.
\newblock {\em\protect\JournalTitle{Soft matter}} \textbf{11}, 6045--6054
  (2015).

\bibitem{siviello2016analysis}
C Siviello, F Greco, D Larobina, Analysis of the aging effects on the
  viscoelasticity of alginate gels.
\newblock {\em\protect\JournalTitle{Soft Matter}} \textbf{12}, 8726--8735
  (2016).

\bibitem{hartley2003logarithmic}
R Hartley, R Behringer, Logarithmic rate dependence of force networks in
  sheared granular materials.
\newblock {\em\protect\JournalTitle{Nature}} \textbf{421}, 928--931 (2003).

\bibitem{brujic2005granular}
J Bruji{\'c}, et~al., Granular dynamics in compaction and stress relaxation.
\newblock {\em\protect\JournalTitle{Physical review letters}} \textbf{95},
  128001 (2005).

\bibitem{siebenburger2012creep}
M Siebenb{\"u}rger, M Ballauff, T Voigtmann, Creep in colloidal glasses.
\newblock {\em\protect\JournalTitle{Physical review letters}} \textbf{108},
  255701 (2012).

\bibitem{suman_analyzing_2019}
K Suman, YM Joshi, Analyzing onset of nonlinearity of a colloidal gel at the
  critical point.
\newblock {\em\protect\JournalTitle{Journal of Rheology}} \textbf{63},
  991--1001 (2019).

\bibitem{hyun2007fourier}
K Hyun, et~al., Fourier-transform rheology under medium amplitude oscillatory
  shear for linear and branched polymer melts.
\newblock {\em\protect\JournalTitle{Journal of Rheology}} \textbf{51},
  1319--1342 (2007).

\bibitem{Berne_Pecora_1976}
B Berne, R Pecora, {\em Dynamic light scattering: with applications to
  chemistry, biology, and physics}.
\newblock (Wiley), (1976).

\bibitem{Cipelletti_Manley_Ball_Weitz_2000}
L Cipelletti, S Manley, RC Ball, DA Weitz, Universal aging features in the
  restructuring of fractal colloidal gels.
\newblock {\em\protect\JournalTitle{Physical Review Letters}} \textbf{84},
  2275–2278 (2000).

\bibitem{madsen_beyond_2010}
A Madsen, RL Leheny, H Guo, M Sprung, O Czakkel, Beyond simple exponential
  correlation functions and equilibrium dynamics in x-ray photon correlation
  spectroscopy.
\newblock {\em\protect\JournalTitle{New Journal of Physics}} \textbf{12},
  055001 (2010).

\bibitem{Lieleg_Kayser_Brambilla_Cipelletti_Bausch_2011}
O Lieleg, J Kayser, G Brambilla, L Cipelletti, AR Bausch, Slow dynamics and
  internal stress relaxation in bundled cytoskeletal networks.
\newblock {\em\protect\JournalTitle{Nature Materials}} \textbf{10}, 236–242
  (2011).

\bibitem{bouzid_elastically_2017}
M Bouzid, J Colombo, LV Barbosa, E Del~Gado, Elastically driven intermittent
  microscopic dynamics in soft solids.
\newblock {\em\protect\JournalTitle{Nature Communications}} \textbf{8}, 15846
  (2017).

\bibitem{tang_anomalous_2015}
S Tang, M Wang, BD Olsen, Anomalous {{Self}}-{{Diffusion}} and {{Sticky Rouse
  Dynamics}} in {{Associative Protein Hydrogels}}.
\newblock {\em\protect\JournalTitle{J. Am. Chem. Soc.}} \textbf{137},
  3946--3957 (2015).

\bibitem{mahmad_rasid_anomalous_2021}
I Mahmad~Rasid, N {Holten-Andersen}, BD Olsen, Anomalous {{Diffusion}} in
  {{Associative Networks}} of {{High}}-{{Sticker}}-{{Density Polymers}}.
\newblock {\em\protect\JournalTitle{Macromolecules}} \textbf{54}, 1354--1365
  (2021).

\bibitem{Bouchaud_Pitard_2001}
JP Bouchaud, E Pitard, Anomalous dynamical light scattering in soft glassy
  gels.
\newblock {\em\protect\JournalTitle{Eur. Phys. J. E}} \textbf{6}, 231–236
  (2001).

\bibitem{bouchaud_anomalous_2008}
JP Bouchaud, Anomalous {{Relaxation}} in {{Complex Systems}}: {{From
  Stretched}} to {{Compressed Exponentials}} in {\em Anomalous {{Transport}}}.
\newblock ({John Wiley \& Sons, Ltd}), pp. 327--345 (2008).

\bibitem{Krall1998}
AH Krall, DA Weitz, Internal dynamics and elasticity of fractal colloidal gels.
\newblock {\em\protect\JournalTitle{Phys. Rev. Lett.}} \textbf{80}, 778 (1998).

\bibitem{Barretta2000}
P Barretta, F Bordi, C Rinaldi, G Paradossi, A dynamic light scattering study
  of hydrogels based on telechelic poly(vinyl alcohol).
\newblock {\em\protect\JournalTitle{The Journal of Physical Chemistry B}}
  \textbf{104}, 11019–11026 (2000).

\bibitem{Usuelli2020}
M Usuelli, et~al., Probing the structure of filamentous nonergodic gels by
  dynamic light scattering.
\newblock {\em\protect\JournalTitle{Macromolecules}} \textbf{53}, 5950–5956
  (2020).

\bibitem{Pommella_Philippe_Phou_Ramos_Cipelletti_2019}
A Pommella, AM Philippe, T Phou, L Ramos, L Cipelletti, Coupling space-resolved
  dynamic light scattering and rheometry to investigate heterogeneous flow and
  nonaffine dynamics in glassy and jammed soft matter.
\newblock {\em\protect\JournalTitle{Physical Review Applied}} \textbf{11},
  034073 (2019).

\bibitem{duri_resolving_2009}
A Duri, DA Sessoms, V Trappe, L Cipelletti, Resolving {{Long}}-{{Range Spatial
  Correlations}} in {{Jammed Colloidal Systems Using Photon Correlation
  Imaging}}.
\newblock {\em\protect\JournalTitle{Phys. Rev. Lett.}} \textbf{102}, 085702--4
  (2009).

\bibitem{evans_introductory_1998}
E Evans, Introductory {{Lecture Energy}} landscapes of biomolecular adhesion
  and receptor anchoring at interfaces explored with dynamic force
  spectroscopy.
\newblock {\em\protect\JournalTitle{Faraday Discuss.}} \textbf{111}, 1--16
  (1998).

\bibitem{lee_direct_2009}
HN Lee, K Paeng, SF Swallen, MD Ediger, Direct {{Measurement}} of {{Molecular
  Mobility}} in {{Actively Deformed Polymer Glasses}}.
\newblock {\em\protect\JournalTitle{Science}} \textbf{323}, 231--234 (2009).

\bibitem{warren_deformation-induced_2010-1}
M Warren, J Rottler, Deformation-induced accelerated dynamics in polymer
  glasses.
\newblock {\em\protect\JournalTitle{The Journal of Chemical Physics}}
  \textbf{133}, 164513 (2010).

\bibitem{sollich_rheology_1997}
P Sollich, F Lequeux, P H{\'e}braud, ME Cates, Rheology of soft glassy
  materials.
\newblock {\em\protect\JournalTitle{Physical review letters}} \textbf{78}, 2020
  (1997).

\bibitem{Duri_Cipelletti_2006}
A Duri, L Cipelletti, Length scale dependence of dynamical heterogeneity in a
  colloidal fractal gel.
\newblock {\em\protect\JournalTitle{Europhysics Letters}} \textbf{76},
  972–978 (2006).

\bibitem{kirch_optical_2012-2}
J Kirch, et~al., Optical tweezers reveal relationship between microstructure
  and nanoparticle penetration of pulmonary mucus.
\newblock {\em\protect\JournalTitle{Proceedings of the National Academy of
  Sciences}} \textbf{109}, 18355--18360 (2012).

\bibitem{ostedgaard_gel-forming_2017}
LS Ostedgaard, et~al., Gel-forming mucins form distinct morphologic structures
  in airways.
\newblock {\em\protect\JournalTitle{Proc Natl Acad Sci USA}} \textbf{114},
  6842--6847 (2017).

\bibitem{zinn2018}
T Zinn, et~al., Ultra-small-angle x-ray photon correlation spectroscopy using
  the eiger detector.
\newblock {\em\protect\JournalTitle{Journal of Synchrotron Radiation}}
  \textbf{25}, 1753--1759 (2018).

\bibitem{macosko1994rheology}
C Macosko, R Larson, K (Firm), {\em Rheology: Principles, Measurements, and
  Applications}, Advances in interfacial engineering series.
\newblock (VCH), (1994).

\bibitem{Truzzolillo2015}
D Truzzolillo, V Roger, C Dupas, S Mora, L Cipelletti, Bulk and interfacial
  stresses in suspensions of soft and hard colloids.
\newblock {\em\protect\JournalTitle{Journal of Physics: Condensed Matter}}
  \textbf{27}, 194103 (2015).

\bibitem{Duri_Bissig_Trappe_Cipelletti_2005}
A Duri, H Bissig, V Trappe, L Cipelletti, Time-resolved-correlation
  measurements of temporally heterogeneous dynamics.
\newblock {\em\protect\JournalTitle{Physical Review E}} \textbf{72}, 051401
  (2005).

\bibitem{pommella2019}
A Pommella, AM Philippe, T Phou, L Ramos, L Cipelletti, Coupling space-resolved
  dynamic light scattering and rheometry to investigate heterogeneous flow and
  nonaffine dynamics in glassy and jammed soft matter.
\newblock {\em\protect\JournalTitle{Phys. Rev. Applied}} \textbf{11}, 034073
  (2019).

\bibitem{duri2009}
A Duri, DA Sessoms, V Trappe, L Cipelletti, Resolving long-range spatial
  correlations in jammed colloidal systems using photon correlation imaging.
\newblock {\em\protect\JournalTitle{Phys. Rev. Lett.}} \textbf{102}, 085702
  (2009).

\bibitem{Cipelletti2013}
L Cipelletti, G Brambilla, S Maccarrone, S Caroff, Simultaneous measurement of
  the microscopic dynamics and the mesoscopic displacement field in soft
  systems by speckle imaging.
\newblock {\em\protect\JournalTitle{Opt. Express}} \textbf{21}, 22353--22366
  (2013).

\bibitem{Cipelletti2003}
L Cipelletti, et~al., Universal non-diffusive slow dynamics in aging soft
  matter.
\newblock {\em\protect\JournalTitle{Faraday Discuss.}} \textbf{123}, 237–251
  (2003).

\bibitem{brenner_laser_1978}
SL Brenner, RA Gelman, R Nossal, Laser {{Light Scattering}} from {{Soft Gels}}.
\newblock {\em\protect\JournalTitle{Macromolecules}} \textbf{11}, 202--207
  (1978).

\bibitem{gelman_laser_1979}
RA Gelman, R Nossal, Laser {{Ligh Scattering}} from {{Mechanically Excited
  Gels}}.
\newblock {\em\protect\JournalTitle{Macromolecules}} \textbf{12}, 311--316
  (1979).

\bibitem{mao_dynamic_1998}
L Mao, H Huglin, Dynamic light scattering from polymer gels: spring-rotor
  model.
\newblock {\em\protect\JournalTitle{Polymer International}} \textbf{45},
  321–326 (1998).

\end{thebibliography}

\newpage
\section*{Supplementary Information}

\subsection*{Mucus rheological characterization}
We report here the results of additional oscillatory tests probing the viscoelastic properties of our mucus samples. Due to the biological origin of mucus, there is a sample-to-sample variability in the magnitude of the viscoelastic moduli, but their frequency and strain amplitude dependence is quite robust. To improve the statistics, we normalize all moduli with respect to the value of $G'$ at small frequency or strain amplitude and present results for the normalized moduli averaged over several independent samples.

\subsubsection*{Frequency Sweeps} 
\begin{figure}
\centering
\includegraphics[width=\columnwidth]{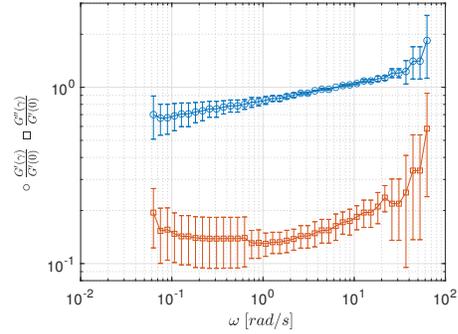}
\caption{Normalized storage (blue circles) and loss (red squares) moduli \textit{vs} frequency \fs{$\omega$} as obtained by averaging at least six frequency sweep tests in the linear regime ($\gamma = 0.01$ ). The bars indicate the standard deviation over the set of samples.}
\label{frequency}
\end{figure}

Figure \ref{frequency} shows the frequency dependence of the viscoelastic moduli, for a strain amplitude $\gamma=0.01$. The normalization was performed by dividing, for each sample, the storage $G^\prime$ and loss $G^{\prime\prime}$ moduli by the magnitude of $G'$  at frequency \fs{$\omega = 6.28 \rs$}. The normalization factors are reported in Table~\ref{value-frequency}.

\newpage

\newpage

\newpage
\subsubsection*{Amplitude Sweeps} 
\begin{figure}
\centering
\includegraphics[width=1\columnwidth]{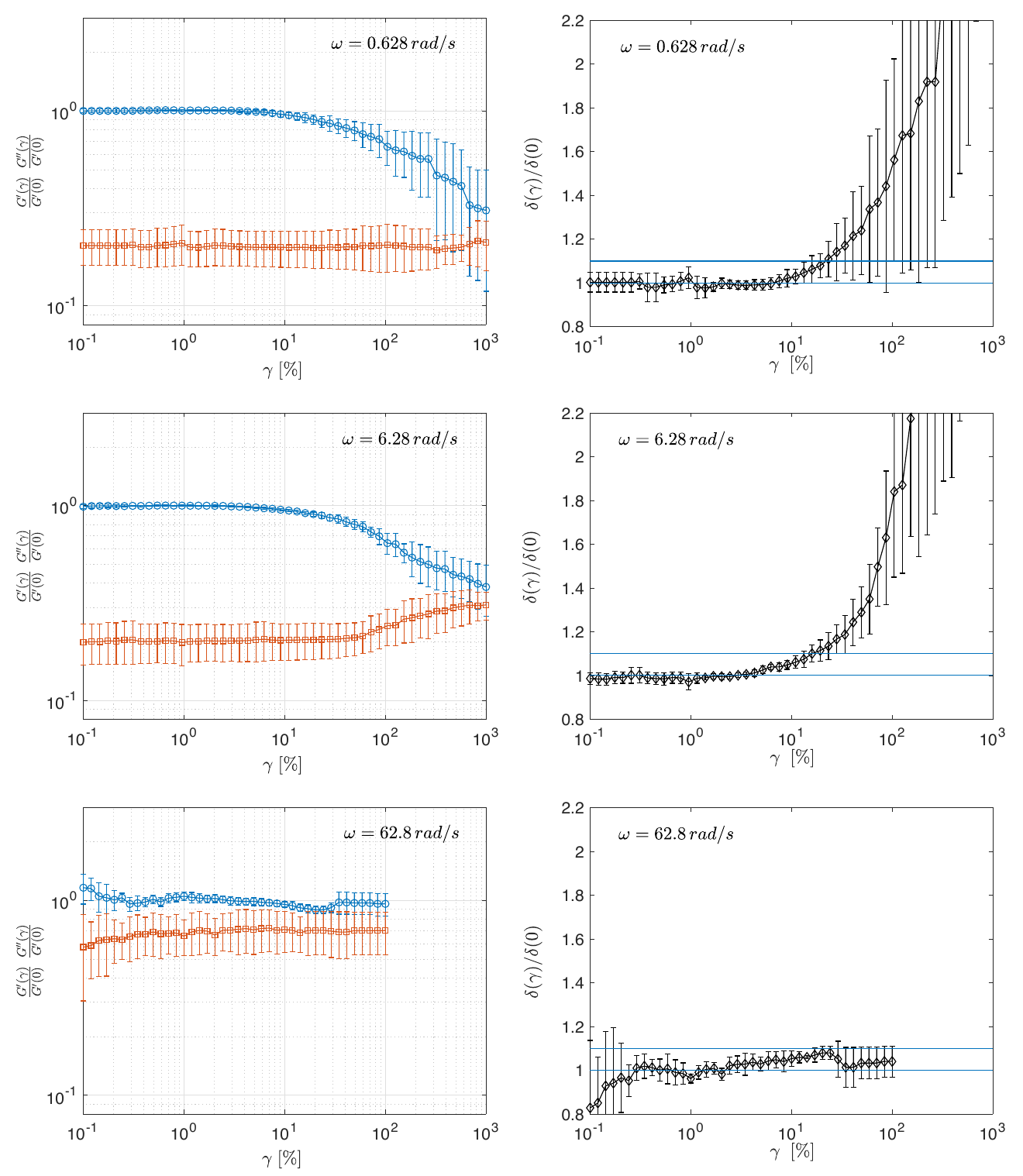}
\caption{Strain amplitude dependence of the viscoelatic moduli in oscillatory tests at various frequencies \fs{$\omega$}. Top, middle, and bottom row: \fs{$\omega = 0.628$, 6.28, and $62.8 \rs$}, respectively. Left column: normalized first harmonic of the storage (blue circles) and loss (red squares) moduli. Right column: corresponding phase shift angle $\delta$ determined from the first harmonic response. The horizontal lines indicate a 10\% deviation from the small $\gamma$ value. In all panels, the bars indicate the standard deviation over all samples. }
\label{fig:amplitude}
\end{figure}

Figure \ref{fig:amplitude} shows the strain-dependent normalized moduli at frequencies \fs{$0.628$, $6.28$, and $62.8$ $\rs$}, respectively. Data for each sample are normalized by dividing the first harmonic of the dynamic moduli, $G_1^\prime(\gamma)$ and $G_1^{\prime\prime}(\gamma)$, by the storage modulus. The latter is taken as the magnitude of $G'_1$ at a strain $\gamma_{min}$ defined as the smallest  in the small-strain regime where the moduli are independent of $\gamma$. Typically, $\gamma_{min} \approx 0.5-1\%$. The normalization factors used in Fig.~\ref{fig:amplitude} are reported in Table~\ref{value-amplitude}: in both cases, we have dropped the subscript 1 to simplify the notation. An analogous procedure is applied to the phase shift angle, $\delta(\gamma) = \tan^{-1}(G''/G')$ evaluated from the same amplitude test.

\newpage

\newpage

\begin{figure}
\centering
\includegraphics[width=\columnwidth]{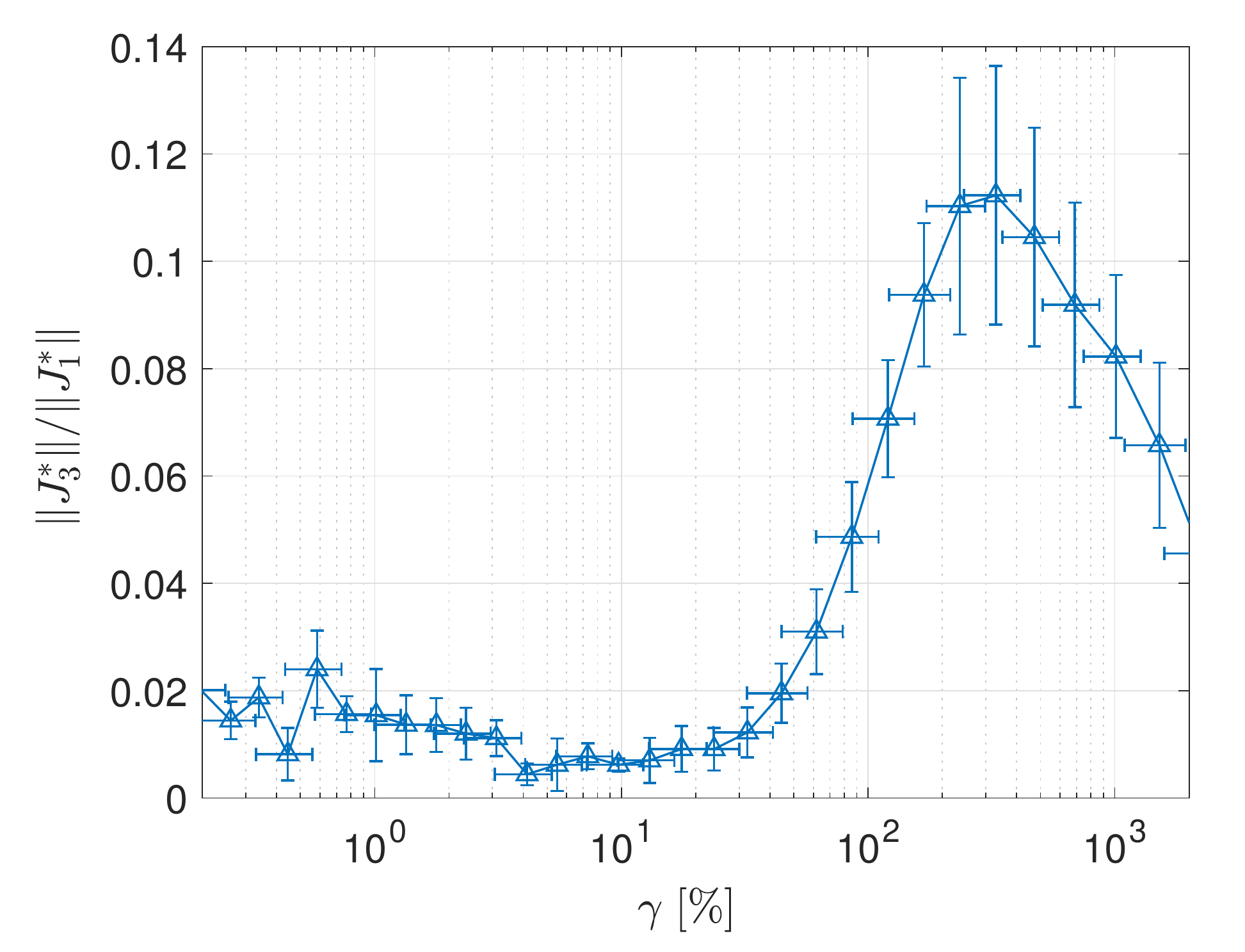}
\caption{Strain dependence of the relative contribution of the norm of the third harmonic of the complex compliance with respect to the first harmonic, both evaluated at \fs{$\omega = 6.28 \rs$}.}
\label{fig:I31}
\end{figure}

Figure~\ref{fig:I31} shows that non-harmonic contributions to the dynamic compliance are negligible up to $\gamma \approx 30\%$, as seen from the initial plateau of the norm of the third harmonic component of the complex compliance $J^*$ normalized by that of the first harmonic.

\subsubsection*{Stress Relaxation} The protocol used for stress relaxation tests consists in measuring the stress evolution for 20 minutes following the application of a given deformation, followed by a rest time of 30 minutes at zero stress before applying subsequent deformations. For the data of Fig 1 of the main manuscript, the applied stress increased monotonically. For the experiments of Fig. 3 of the main manuscript, the duration of the shear ramp at $\dot{\gamma} = 0.001, 0.005, 0.01, 0.02$, and 0.04 s$^{-1}$ was $200, 40, 20, 10$, and 5 s, respectively, so as to achieve the same deformation $\gamma_0 = 20\%$ at the end of each ramp. The shear ramps where performed in the order $\dot{\gamma} = 0.001, 0.01, 0.02, 0.04$, and 0.005 s$^{-1}$, leaving the sample at rest 30 minutes after each ramp.

\newpage
\newpage

\subsubsection*{Tracer particles for XPCS do not modify the rheological properties of mucus}
\begin{figure}
\centering
\includegraphics[width=\columnwidth]{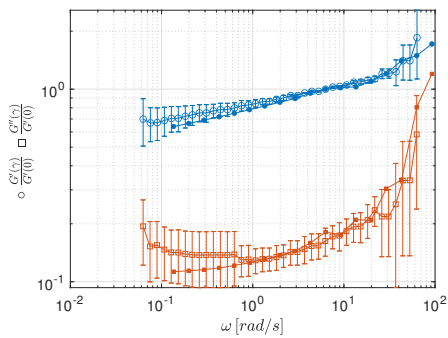}
\caption{\fs{Frequency dependence of the} normalized storage and loss moduli of raw mucus (open symbols, same data as in Fig.~\ref{frequency}) and of mucus loaded with $1.25\%$ w/w silica nanoparticles of radius $\approx 18$ nm (filled symbols). For the sample with nanoparticles, $G^{\prime} = 16.13$ Pa}
\label{fig:rheosilica}
\end{figure}

In the stress relaxation test simultaneous to XPCS measurements (Fig. 2b and 2c of the main manuscript), we added 1.25\% w/w of silica nanoparticles (Ludox TM50 by Aldrich, radius $\approx 18$ nm), to enhance the scattering signal. Figure~\ref{fig:rheosilica} shows that the addition of such a small amount of nanoparticles does not modify the viscoelastic properties of mucus.

\newpage
\newpage


\subsection*{Measurements of the mucus microscopic dynamics}

\subsubsection*{Rheo-XPCS setup} 
\begin{figure}[h]
\centering
\includegraphics[width=\columnwidth]{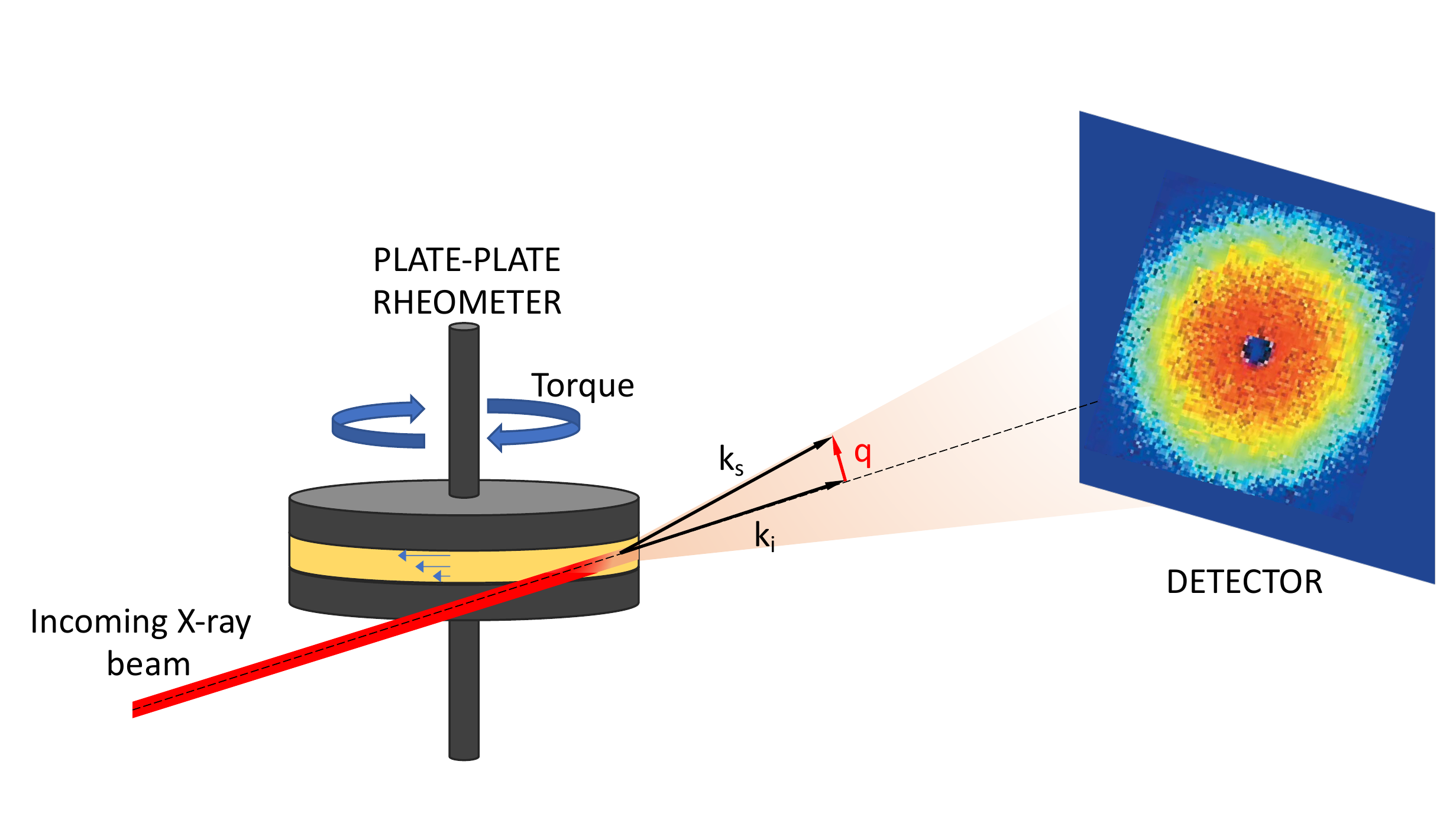}
\caption{Schematic view of the ultra-small angle X-ray photon correlation spectroscopy setup coupled to a stress-controlled rheometer. $\mathbf{k_i}$ and $\mathbf{k_s}$ are the wave vectors of the incoming and scattered beams, respectively; $\mathbf{q}$ is the scattering vector. The  beam illuminates the sample at mid height between the plates, at a radial distance $r=9$ mm from the toll axis, about 1 mm from the plate edge.}
\label{rheo-xpcs}
\end{figure}

The setup used at the ESRF ID-02 line for simultaneous XPCS and rheological measurements is schematically shown in Fig.~\ref{rheo-xpcs}, see Ref.~\cite{zinn2018} for more details. The X-ray beam crosses the sample at half height of the plates gap ($e=1$ mm) entering parallel to the plate surfaces in a tangential geometry. In this configuration, when a step strain is imposed by the rheometer the beam illuminates a sample portion subjected to a uniform shear equal to approximately 3/2 of the nominal strain reported by the rheometer. The detector was a Eiger-500K pixel array detector. Two-time intensity correlation functions were calculated according to
\begin{equation} \label{eq:g2_xpcs}
g_2 (t, \tau) - 1 = B \frac{\langle I_p(t)I_p(t+\tau) \rangle_{\varphi}} {\langle I_p(t) \rangle_{\varphi}\langle I_p(t+\tau) \rangle_{\varphi}} - 1 \,,
\end{equation}
with $I_p(t)$ the time-dependent intensity of the $p$-th pixel,  $t$ the time after imposing a strain step, $\tau$ a time delay, and $ < \cdot  \cdot  \cdot >_{\varphi}$ the azimuthal average over a ring of pixels centered around the transmitted beam position. \resub{The prefactor $B$ is chosen such that $g_2 (t, \tau) \rightarrow 1$ for $\tau \rightarrow 0$ and is obtained by fitting the initial decay of $\ln(g_2-1)$ to a polynomial of order two.}

\newpage
\newpage

\subsubsection*{Rheo-DLS setup} 
\begin{figure*}
\centering
\includegraphics[width=\columnwidth]{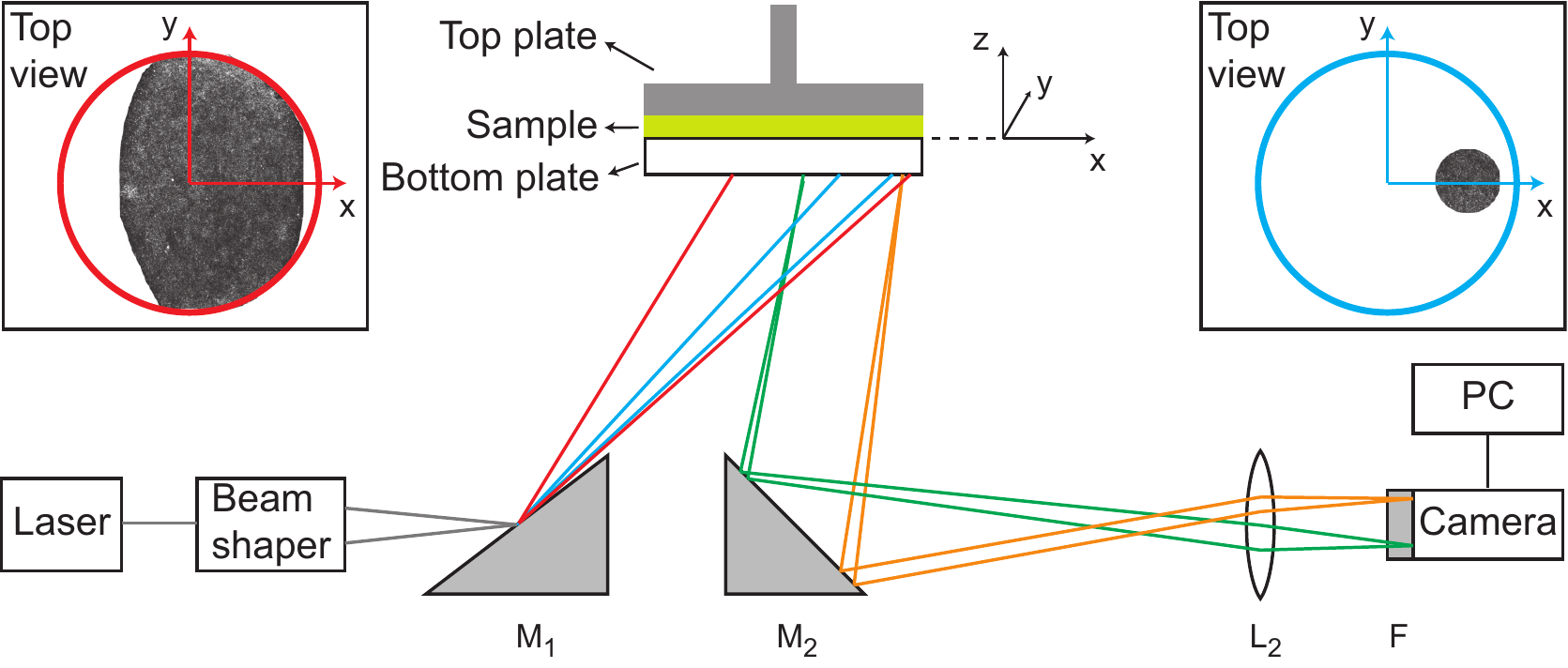}
\caption{Scheme of the wide-angle dynamic light scattering setup coupled to a stress-controlled rheometer. The incoming beam (gray rays) is sent to the sample by the mirror $M_1$ and can be adjusted to illuminate almost the entire sample (red rays and left top view), or just a small sample area (blue rays and right top view). The back-scattered light (orange and green lines) is imaged on the camera using the mirror $M_2$ and the objective lens $L_2$. The $x-y$ plane coincides with the interface between the sample and the bottom plate.}
\label{rheo-dls}
\end{figure*}

The home-made wide-angle DLS setup for coupled rheology and light scattering measurements is shown in Fig~\ref{rheo-dls} (see Ref.~\cite{pommella2019} for a detailed description of the apparatus). Stress relaxation tests and tests at constant shear rate were performed in the plate-plate geometry using a transparent glass window as the bottom plate. The incoming laser beam is sent to the sample by the mirror $M_1$. The beam size can be adjusted to illuminate almost the entire sample, as indicated by the red rays and the top view on the left, or a small sample area of linear size approximately 10 mm, at 2/3 of the plate radius, as indicated by the blue rays and in the top view on the right. The back-scattered light (orange and green rays) is collected by the mirror $M_2$ and imaged on the detector of a CMOS camera through the objective lens $L_2$,  to allow for space-resolved measurements. In this configuration, each pixel of the camera corresponds to a specific area of the sample in the $x-y$ plane defined as the interface between the sample and the bottom plate. The scattering vector $q$ of the setup is fixed and equal to $q=33$ $\mu$m$^{-1}$, which corresponds to a characteristic length scale $\sim \frac{\pi}{q} = 95$ nm.

The microscopic dynamics are probed in a space- and time-resolved manner by acquiring with a CMOS camera a time series of speckle images, at a rate of 0.09 Hz. In order to investigate non-stationary dynamics, the images are processed to calculate a space-resolved, two-time intensity autocorrelation function \cite{duri2009, Cipelletti2013}:
\begin{equation} \label{eq:g2}
g_2 (t, \tau, r) - 1 = B \frac{\langle I_p(t)I_p(t+\tau) \rangle_{r}} {\langle I_p(t) \rangle_{r}\langle I_p(t+\tau) \rangle_{r}} - 1
\end{equation}
Here, $I_p(t)$ is the time-dependent intensity of the $p$-th pixel, $t$ the time after reaching the target strain, $\tau$ a time delay. $ < \cdot  \cdot  \cdot >_{r}$ indicates an average over a ring of pixels with thickness 450 $\mu$m and distance $r$ from the rheometer axis, corresponding to a portion of the sample with the same stress and strain in the plate-plate geometry. \resub{The prefactor $B$ is chosen such that $g_2 (t, \tau) \rightarrow 1$ for $\tau \rightarrow 0$ and is obtained by fitting the initial decay of $\ln(g_2-1)$ to a polynomial of order two.}

\newpage

\newpage
\subsection*{Model}

\subsubsection*{Parameters $A$ and $\beta$ for known dynamics} In deriving the model, we used
\begin{equation}
    g_2\left [ \Delta r (n,\delta \bar r),q\right]-1 = \exp[-A(q n^p\delta \bar r)^\beta] \,.
    \label{eq:g2general}
\end{equation}
We show here that this expression reduces to known forms of $g_2-1$ in the limiting cases of diffusive or ballistic dynamics. For diffusive dynamics, $p=0.5$ and $\Delta r^2 = n \delta \bar r^2$, with $n$ the number of steps of length $\delta \bar r$ of a random walk. Equation~\ref{eq:g2general} with $A=1/3$ and $\beta=2$ then reduces to $g_2-1 = \exp(-\Delta r^2 q^2 /3) = \exp(-2 D q^2 \tau)$, which is the usual expression for $g_2-1$ for particles with diffusion coefficient $D= \Delta r^2/6\tau$~\cite{Berne_Pecora_1976}. For ballistic dynamics, $p=1$ and $\Delta r = n \delta \bar r$, such that Eq.~\ref{eq:g2general} with $A=2$ becomes $g_2-1 = \exp[-2(q\Delta r)^\beta] = \exp[-2(q V_0 \tau)^\beta]$. The latter is the form predicted for ballistic dynamics resulting from the relaxation of internal stress, with $V_0$ the typical velocity of the ballistic motion and $\beta > 1$ an exponent characterizing the Levy probability distribution of the velocities~\cite{Cipelletti2003}.

\subsubsection*{Fast relaxation and spontaneous dynamics} 
\begin{figure*}
\centering
\includegraphics[width=\columnwidth]{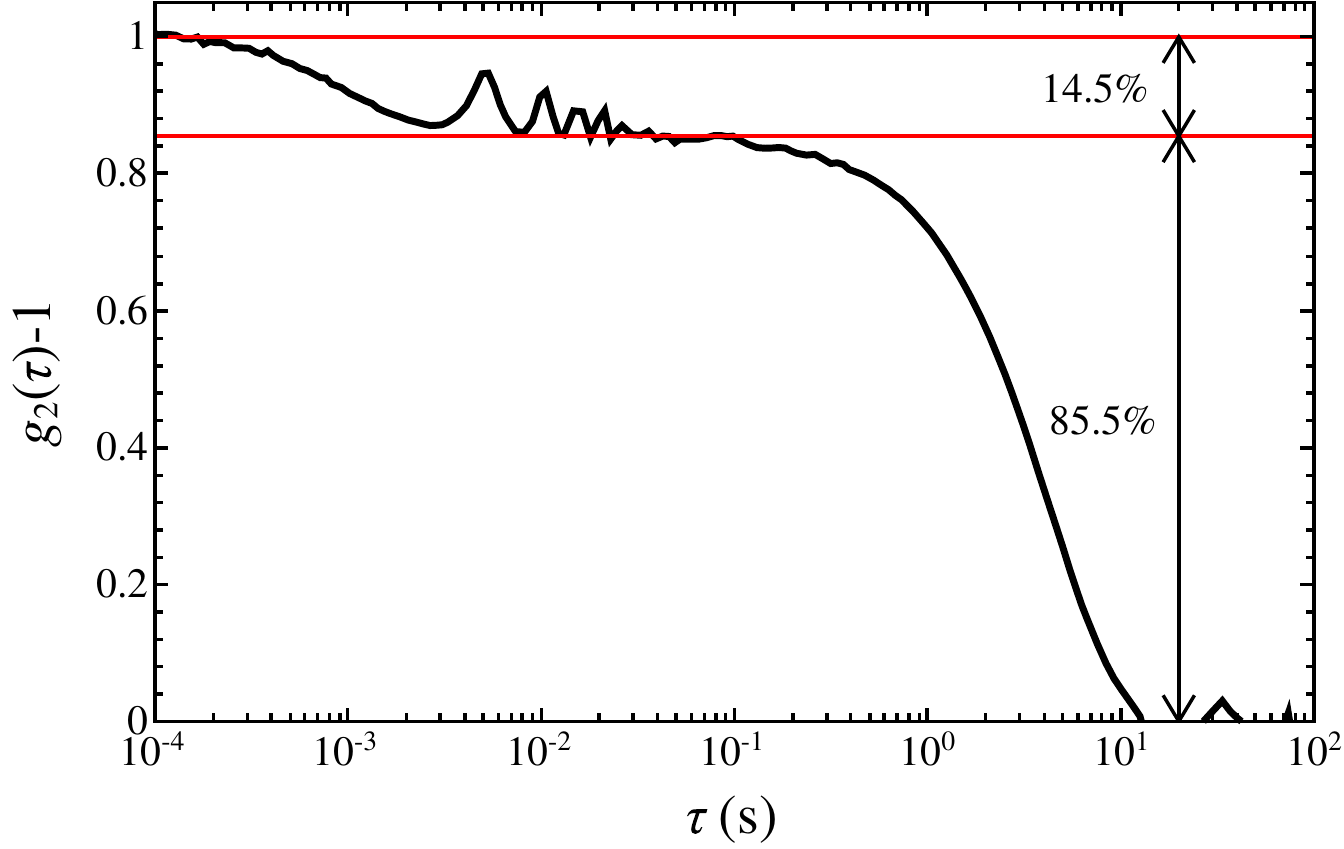}
\caption{\resub{Fast dynamics of a mucus gel at rest as measured by conventional dynamic light scattering in a setup with $\lambda = 640$ nm, scattering angle $\theta = 165$ deg and scattering vector $q = 25.9~\mu\mathrm{m}^{-1}$. The fast decay accounts for less than 15\% of the full relaxation of $g_2-1$.}}
\label{dls-fast}
\end{figure*}

The model does not account for the time it takes the network to relax to a new configuration of mechanical equilibrium after a bond breaking. This time is comparable to that of thermally activated fluctuations of the network at fixed connectivity\resub{, corresponding to the fastest relaxation mode of the gel. Figure~\ref{dls-fast} shows the intensity correlation function measured by conventional DLS on a mucus gel at rest. The fast dynamics have a time scale $\lesssim 1$ ms, similar to that of other colloidal and polymeric gels~\cite{Krall1998,Barretta2000,Usuelli2020}. The fast mode accounts for less than 15\% of the full relaxation of the intensity correlation function $g_2-1$. The oscillations in the range 5-20 ms are due to ringing modes, which are easily excited by even slight mechanical disturbances and are very often seen in polymer gels (see e.g.~\cite{brenner_laser_1978,gelman_laser_1979,mao_dynamic_1998}).} Since the shortest time delay probed by our experiments is 50 ms, these fast dynamics may be safely neglected. The model furthermore neglects the underlying spontaneous slow dynamics that occurs even in the absence of an applied stress (see Fig.~2a of the main manuscript). Accordingly, in Figs.~3-5 of the main manuscript we restrict the analysis of the dynamics to the accelerated regime where the relaxation rate \resub{of the dynamics associated with the relaxation of the externally imposed stress} is significantly larger than that of the spontaneous dynamics. \resub{More specifically, for XPCS we include data for $t \le 3$ s, while for the DLS experiments of Figs. 3 and 4 of the main manuscript we include all data up to the first time the decay rate $\Gamma$ decreases below $0.09~\mathrm{s}^{-1}$. In Figs. 5d,e of the main manuscript the range of data for which the scaling is tested corresponds to $0.1 \le g_2-1 \le 0.99$.}

\newpage

\newpage

\begin{table}
\centering
\caption{Values of the normalizing storage modulus used in the frequency sweep tests}

\begin{tabular}{lrrr}
Frequency (\fs{$\rs$}) & \fs{6.28} \\
\midrule

 G' (Pa)    & 17.00 \\
            &  24.18 \\
            &  25.44 \\
            &  28.64 \\
            &  32.28 \\
            &  33.07 \\
            &  43.97 \\
            &  47.04 \\
            &  55.61 \\
            &  96.90 \\
            &  185.50 \\
\bottomrule
\end{tabular}
\label{value-frequency}
\end{table}

\newpage

\begin{table}
\centering
\caption{Values of the normalizing storage modulus used in the amplitude sweep tests, Fig.~\ref{fig:amplitude}}

\begin{tabular}{lrrr}
Frequency \fs{($\rs$)} & \fs{0.628} & \fs{6.28} & \fs{62.8} \\
\midrule

G' (Pa) & 8.1043 &  19.0613 &  20.4418 \\
        &  10.0298 &  21.1750 &  21.4767 \\
        &  13.5550 & 122.1333 &  20.0000 \\
        &  35.0580 & 112.9833 &  35.7075 \\
        &  23.9700 &  49.1960 &  39.1667 \\
        &  30.7450 &  35.6560 &         \\
        &         &  23.9050 &         \\
        &         &  31.0150 &         \\
\bottomrule
\end{tabular}
\label{value-amplitude}
\end{table}
\newpage







\end{document}